\documentclass[twocolumn]{aastex63}

\received{June, 2022}
\revised{--}
\accepted{--}

\submitjournal{ApJ}

\shorttitle{IMACS spectroscopy of Grus I}
\shortauthors{Chiti et al.}

\newcommand{\kms}         {km\,s$^{-1}$}

\newcommand{\masyr}       {mas~yr$^{-1}$}
\newcommand{\mua}         {\mu_{\alpha}\cos{\delta}}
\newcommand{\mud}         {\mu_{\delta}}

\def\spose#1{\hbox to 0pt{#1\hss}}
\def\lta{\mathrel{\spose{\lower 3pt\hbox{$\mathchar"218$}}
     \raise 2.0pt\hbox{$\mathchar"13C$}}}
\def\gta{\mathrel{\spose{\lower 3pt\hbox{$\mathchar"218$}}
    \raise 2.0pt\hbox{$\mathchar"13E$}}}

\begin{document}

\title{Magellan/IMACS spectroscopy of Grus I: A low metallicity ultra-faint dwarf galaxy\footnote{This paper includes data gathered with the 6.5 meter Magellan Telescopes located at Las Campanas Observatory, Chile.}}

\correspondingauthor{Anirudh Chiti}
\email{achiti@uchicago.edu}

\author[0000-0002-7155-679X]{Anirudh Chiti}
\affil{Department of Astronomy \& Astrophysics, University of Chicago, 5640 S Ellis Avenue, Chicago, IL 60637, USA}
\affil{Kavli Institute for Cosmological Physics, University of Chicago, Chicago, IL 60637, USA}

\author[0000-0002-4733-4994]{Joshua D. Simon}
\affiliation{Observatories of the Carnegie Institution for Science, 813 Santa Barbara St., Pasadena, CA 91101, USA}

\author[0000-0002-2139-7145]{Anna Frebel}
\affiliation{Department of Physics and Kavli Institute for Astrophysics and Space Research, Massachusetts Institute of Technology, Cambridge, MA 02139, USA}

\author[0000-0002-6021-8760]{Andrew B. Pace}
\affiliation{McWilliams Center for Cosmology, Carnegie Mellon University, 5000 Forbes Ave, Pittsburgh, PA 15213, USA }

\author[0000-0002-4863-8842]{Alexander P. Ji}
\affil{Department of Astronomy \& Astrophysics, University of Chicago, 5640 S Ellis Avenue, Chicago, IL 60637, USA}
\affil{Kavli Institute for Cosmological Physics, University of Chicago, Chicago, IL 60637, USA}

\author[0000-0002-9110-6163]{Ting S. Li}
\affiliation{Department of Astronomy and Astrophysics, University of Toronto, 50 St. George Street, Toronto ON, M5S 3H4, Canada}

\begin{abstract}

We present a chemodynamical study of the Grus I ultra-faint dwarf galaxy (UFD) from medium-resolution ($R\sim11,000$) Magellan/IMACS spectra of its individual member stars. 
We identify eight confirmed members of Grus~I, based on their low metallicities and coherent radial velocities, and four candidate members for which only velocities are derived.
In contrast to previous work, we find that Grus~I has a very low mean metallicity of $\langle$[Fe/H]$\rangle = -2.62 \pm 0.11$\,dex, making it one of the most metal-poor UFDs.
Grus~I has a systemic radial velocity of $-143.5\pm1.2$\,km\,s$^{-1}$ and a velocity dispersion of $\sigma_{\text{rv}} = 2.5^{+1.3}_{-0.8}$\,km\,s$^{-1}$ which results in a dynamical mass of $M_{1/2}\,(r_h) = 8^{+12}_{-4} \times 10^5$\,M$_{\odot}$ and a mass-to-light ratio of M/L$_V$ = $440^{+650}_{-250}$\,M$_\odot$/L$_\odot$.
Under the assumption of dynamical equilibrium, our analysis confirms that Grus~I is a dark-matter-dominated UFD (M/L $> 80$\,M$_\odot$/L$_\odot$).
However, we do not resolve a metallicity dispersion ($\sigma_{\text{[Fe/H]}} < 0.44$\,dex).
Our results indicate that Grus~I is a fairly typical UFD with parameters that agree with mass-metallicity and metallicity-luminosity trends for faint galaxies.
This agreement suggests that Grus~I has not lost an especially significant amount of mass from tidal encounters with the Milky Way, in line with its orbital parameters. 
Intriguingly, Grus~I has among the lowest central density ($\rho_{1/2} \sim 3.5_{-2.1}^{+5.7} \times 10^7$\,M$_\odot$\,kpc$^{-3}$) of the UFDs that are not known to be tidally disrupting.
Models of the formation and evolution of UFDs will need to explain the diversity of these central densities, in addition to any diversity in the outer regions of these relic galaxies. 

\end{abstract}


\section{Introduction} 
\label{sec:intro}

Over the last two decades, data from large digital sky surveys (e.g., the Sloan Digital Sky Survey, the Dark Energy Survey, PanSTARRS) have led to an order-of-magnitude increase in the number of known low surface brightness stellar systems in the vicinity ($< 200$\,kpc) of the Milky Way \citep[e.g.,][]{wbd+05, wdm+05, zbe+06, bze+07, wjw+07, w+10, bdb+15, dbr+15, kj+15, kjm+15, kbt+15, lmi+15, lmb+15, dba+16, hco+16, hco+18, mcp+20, cpd+21, csl+22}.
One particularly intriguing class of faint systems, which were first detected in SDSS data  over a decade ago \citep{wbd+05, wdm+05}, are ultra-faint dwarf galaxies (UFDs).
UFDs are the least luminous ($L < 10^5$\,L$_{\odot}$; \citealt{s+19}), most dark-matter-dominated ($\gtrsim 100$\,M$_\odot$/L$_\odot$; \citealt{sg+07}), and among the oldest ($\sim13$\,Gyr; e.g., \citealt{btg+14}) stellar systems. 
Consequently, they are unique nearby probes of galaxy formation on the smallest scales \citep[e.g.][]{rpa+19}, early chemical evolution \citep[e.g.][]{fsk+14}, and indirect signatures of dark matter interactions \citep[e.g.][]{hess+20}. 

An initial step in characterizing UFDs is to derive their dynamical masses and metallicities through spectroscopy of member stars.
This step is crucial, because it is often ambiguous upon discovery whether a faint system is a bona-fide UFD or a globular cluster (GC) \citep{ws+12}.
One distinguishing feature between GCs and UFDs is that GCs are not dark-matter-dominated (M/L $\lesssim 3$\,M$_\odot$/L$_\odot$; \citealt{mv+05}) whereas UFDs are the most dark-matter-dominated known systems ($\gtrsim 100$\,M$_\odot$/L$_\odot$; \citealt{sg+07}).
From spectroscopy, one can derive the velocity dispersion of member stars, which can then be used to derive a dynamical mass \citep{wmo+09, wmb+10} and an accompanying mass-to-light ratio for classification.
A metallicity dispersion can also be used to separate GCs from UFDs, since UFDs show significant metallicity spreads \citep[e.g.,][]{fsk+14} whereas GCs show minimal spreads ($\lesssim 0.05$\,dex; e.g., \citealt{cbg+09}). 
Differences in chemical abundance patterns are also observed \citep[e.g.,][]{jsf+19}.

Simply identifying a system as a UFD and presenting its general properties (e.g., dynamical mass, mean metallicity \& dispersion) is of scientific interest.
For instance, the number and distribution of the Milky Way's UFDs can constrain models of dark matter \citep[e.g.][]{kph+18, ndb+21, mnw+22}.
Whether or not a UFD lies on the luminosity-mass relation for dwarf galaxies \citep{kcg+13} can indicate mass loss from interactions with the Milky Way \citep[e.g.,][]{sld+17, lsk+18}.
And the kinematics of stars in a UFD can be used to derive a J-factor to determine the relevance of the system for indirect searches for signatures of dark matter interactions \citep[e.g.,][]{ps+19}.

In this regard, Grus~I is a particularly interesting Milky Way satellite as its classification has remained somewhat ambiguous.
The system was first discovered in Dark Energy Survey (DES) DR1 data as a metal-poor, faint ($M_V = -3.4 \pm 0.3$) satellite \citep{kbt+15}; although, its location near a DECam chip-gap in the DES imaging rendered its structural properties uncertain.
Subsequent, wide-field photometric follow-up \citep{cpm+21} derived a half-light radius of $r_h = 151^{+21}_{-31}$\,pc and a photometric $\langle$[Fe/H]$\rangle = -1.88^{+0.09}_{-0.03}$, suggesting Grus I to be a UFD due to its large size but with a higher than typical UFD metallicity.
\citet{jck+18} found a lower metallicity of $\langle$[Fe/H]$\rangle = -2.50^{+0.30}_{-0.30}$ in their photometric study of Grus~I.
An initial spectroscopic study of Grus I by \citet{wmo+16} was unable to resolve a velocity dispersion and derived a mean metallicity of $\langle$[Fe/H]$\rangle = -1.42^{+0.55}_{-0.42}$ from seven probable member stars. 
Such a metallicity, at face value, is abnormally high for a UFD (typically $\langle$[Fe/H]$\rangle < -2.0$; \citealt{s+19}), but the large uncertainty on the metallicity precluded a clear classification.

Recent observations have hinted that Grus I is a UFD.
The brightest two stars in Grus I have [Fe/H]~$\approx-2.5$ and show deficiencies in neutron-capture element abundances, which is a distinctive signature of UFD stars \citep{jsf+19}. 
A recent study presented in \citet{zjb+21} supports the classification of Grus I as a UFD by detecting a large, but uncertain, velocity dispersion of $10.4^{+9.3}_{-5.1}$\,km\,s$^{-1}$ using MUSE data ($R\sim3000$). 
However, this study does not derive spectroscopic metallicities for its sample of members, and only selects Grus~I member stars through a kinematic selection. 
This can artificially inflate the derived velocity dispersion by making the sample susceptible to contamination from more metal-rich foreground Milky Way halo stars that have similar kinematics to Grus I.
A detailed and uniform study of Grus~I, with a joint metallicity and kinematic analysis, is therefore needed to conclusively determine its nature.

In this paper, we present a comprehensive study of member stars in Grus I with new spectroscopy from Magellan/IMACS.  
We re-observe all likely members that were presented in \citet{wmo+16} to derive independent velocity and metallicity measurements, and to search for any binary stars. 
From our joint metallicity and velocity analysis, we identify eight member stars and derive a mean metallicity of $\langle$[Fe/H]$\rangle = -2.62 \pm 0.11$ and a velocity dispersion of $\sigma = 2.5^{+1.3}_{-0.8}$\,km\,s$^{-1}$, confirming that Grus I is a canonical dark-matter-dominated UFD (M/L$_V$ = $440^{+650}_{-250}$\,M$_{\odot}$/L$_{\odot}$ and M/L$_V$ $> 80$\,M$_{\odot}$/L$_{\odot}$). 
We then comment on the orbital history and evolution of Grus I, based on its derived properties and our sample of members.

The paper is organized as follows. In Section~\ref{sec:obs}, we describe our observations; In Section~\ref{sec:analysis}, we outline our methodology in deriving velocities, metallicities, and identifying members; In Section~\ref{sec:discussion}, we derive the dynamical mass, mean metallicity, and orbit of Grus I, and comment on its evolution; and in Section~\ref{sec:conclusion}, we conclude.

\section{Observations \& Data Reduction}
\label{sec:obs}

\begin{deluxetable*}{lllllcllll} 
\tablecolumns{8}
\tablecaption{\label{tab:obs} Observations}
\tablehead{   
  \colhead{Mask} &
  \colhead{RA (h:m:s)} & 
  \colhead{DEC (d:m:s)} &
  \colhead{Slit PA} &
  \colhead{$t_{\text{exp}}$} &
  \colhead{Date of Observation} &
  \colhead{MJD of} &
  \colhead{Number of } & 
  \colhead{Number of } \\
   Name &
  (J2000) &
  (J2000) & 
  (deg) &
  (min) &
  (MM/DD/YYYY) &
  Observation\tablenotemark{a} &
  slits &
  useful spectra\tablenotemark{b}
}
\startdata
Mask 1 & 22:56:40 & $-50$:10:40 & 90 & 225 & 07/26/2015 & 57229.33 & 28 & 17  \\
Mask 2 & 22:56:36 & $-50$:08:59 & 214 & 180 & 10/06/2019 & 58762.03 & 72 & 61  \\
Mask 3\tablenotemark{c} & 22:56:53 & $-50$:08:30 & 72 & 30 & 09/14/2021 & 59471.02 & 6 & 3  \\
\enddata
\tablenotetext{a}{For masks observed over multiple nights, we list the midpoint MJD of observation.}
\tablenotetext{b}{Defined as having a velocity measurement in Table~\ref{tab:gru1spec}.}
\tablenotetext{c}{Mask 3 was designed to obtain additional velocities for previously confirmed, bright Grus I members in Masks 1 and 2.}

\end{deluxetable*}

\subsection{Summary of Observations}

We observed Grus I using the IMACS spectrograph \citep{dhb+06} on the Magellan-Baade Telescope with three separate multi-slit masks in July 2015, October 2019, and September 2021, respectively (see Table~\ref{tab:obs} for details).
We operated the spectrograph following previous multi-slit spectroscopic studies of UFDs using the IMACS instrument \citep[e.g.][]{sld+17, lsd+17, sle+20}, which we briefly outline here.
The observations were performed using the f/4 camera, which nominally granted a 15\farcm4 by 15\farcm4 field of view for slit placement.
We used a slit size of 0\farcs7 and the 1200\,$\ell$\,mm$^{-1}$ grating at a tilt angle of 32\fdg4, which granted a resolution of $R\approx11,000$ and wavelength range of $\sim7500$\,{\AA} to $\sim9000$\,\AA.
This wavelength range is sufficient to cover the prominent telluric A-band feature ($\sim7600$\,\AA) and the calcium triplet absorption lines (8498\,\AA, 8542\,\AA, 8662\,\AA).
This setup grants a minimum velocity precision of $\sim1$\,km\,s$^{-1}$ \citep[e.g.,][]{sld+17}.
We note that the exact wavelength range varies for each spectrum based on the location of the slit on the multi-slit mask.
However, we placed slits to ensure that at least the calcium triplet region (8450\,{\AA} to 8700\,{\AA}) was covered for each star.

Our observing sequence included 2-3 science exposures of 1800\,s to 3300\,s, followed by an arc frame for wavelength calibration, and then a quartz frame for order-tracing and flat-fielding purposes.
We note that we used HeNeAr reference lamps for the wavelength calibration of our observations in 2015, but switched to KrHeNeAr lamps for all subsequent observations in order to make use of strong Kr lines between 7600\,{\AA} and 7900\,\AA. 
The weather was mediocre ($\sim1$\farcs0 seeing) during the 2015 observations, $\sim0\farcs7$ seeing with occasional cirrus for the 2019 observations, and $\sim0\farcs6$ seeing with clear skies for the 2021 observations.
Table~\ref{tab:obs} lists the details of our observations.

We reduced the data following \citet{sld+17} and \citet{lsd+17}.
The COSMOS reduction pipeline \citep{dbh+11, ock+17} was used to locate the slits on the CCD array, generate an initial wavelength solution, and extract 2D spectra.
Then, we used a modified version of the DEEP2 reduction pipeline \citep{cnd+12, ncd+13} that had been altered by \citet{sld+17} to refine the IMACS wavelength solution and extract 1d spectra. 

\subsection{Target Selection}

We observed Grus~I using three multi-slit masks (see Table~\ref{tab:obs}). 
Targets for Mask 1 were selected by overlaying a 12\,Gyr, [Fe/H] = $-2.5$ Dartmouth isochrone \citep{dcj+08} at the assumed distance modulus of Grus~I  ($m - M = 20.4$; \citealt{kbt+15}) on a $g, r$ color-magnitude diagram of stars within 20\farcm0 of Grus~I.
The color-magnitude diagram for this selection was generated by running a default configuration of Source Extractor \citep{ba+96} on DES images of Grus I that were retrieved from the NOAO public data archive \citep{fdh+15, mgm+18}. 
We identified stars within 0.1\,mag of the isochrone as candidate Grus~I members and ultimately selected 28 stars for inclusion on this mask, limited by constraints arising from slit placement.

Mask 2 was designed using photometry from DES DR1 \citep{abbott18}.  Target selection was carried out with a 12.5~Gyr, [Fe/H] = $-2.3$ Dartmouth isochrone, identifying stars within 0.08~mag of the red giant branch (RGB) and with Gaia DR2 proper motions consistent with that of Gru~I.  Eleven RGB candidates were included on the mask.  Mask 3 was designed to obtain an additional measurement of the two brightest member stars identified from the previous masks, which each exhibited possible signs of weak radial velocity variations.

\section{Analysis}
\label{sec:analysis}

\begin{figure*}[!htbp]
\centering
\includegraphics[width =\textwidth]{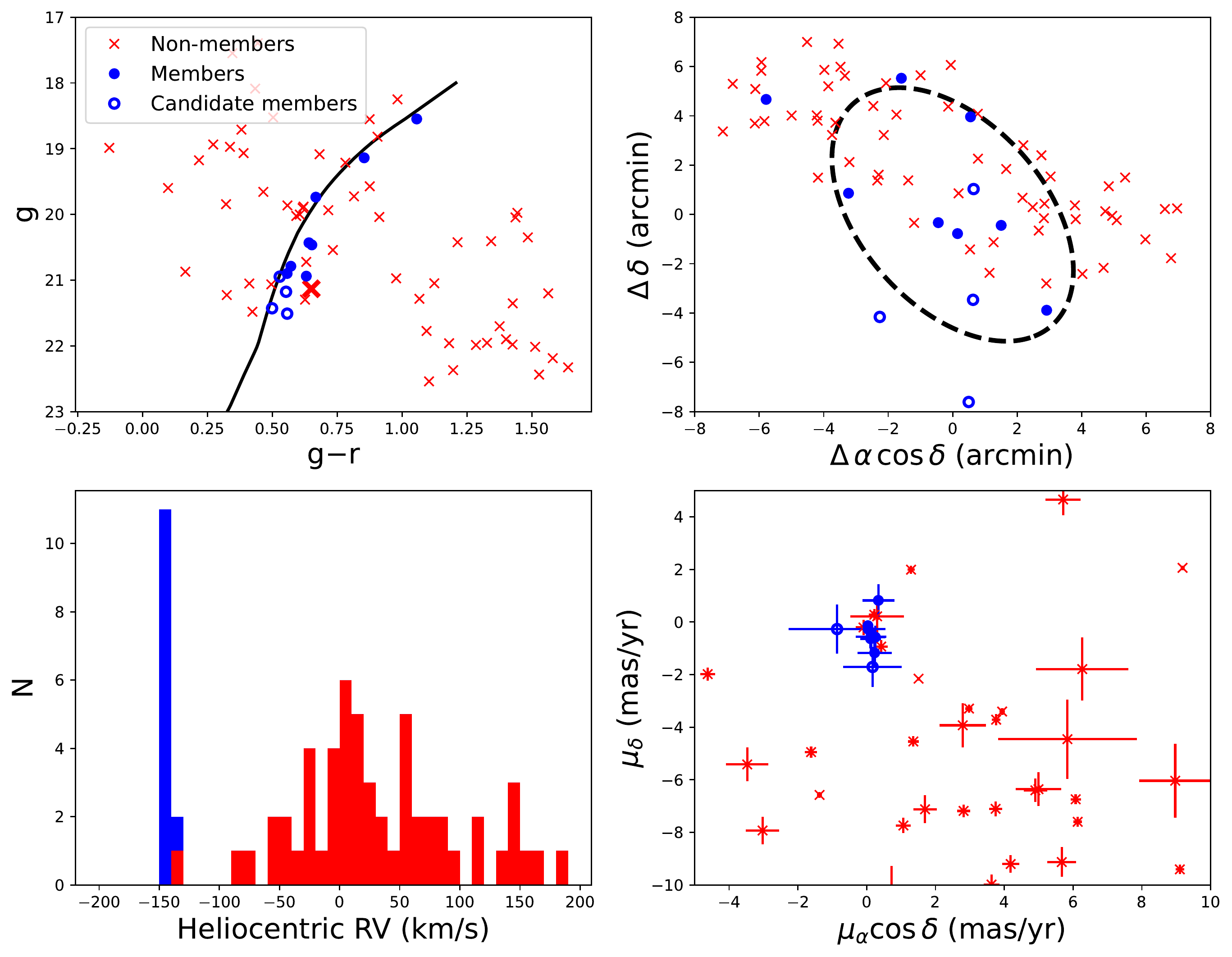}
\caption{Top left: Color-magnitude diagram of stars observed in this study. Confirmed non-members of Gru I, as determined by their radial velocities and metallicities (see Section~\ref{sec:membership}) are shown as red crosses.
Confirmed members of Grus I are shown as filled blue circles.
Candidate members (those with velocities consistent with membership, but no metallicity information) are shown as hollow blue circles. 
DES\,J225643.20$-$501130.0, the star with a radial velocity consistent with Grus~I membership but a high metallicity ([Fe/H] $= -0.9 \pm 0.38$; see Section~\ref{sec:membership}) is indicated by a larger red cross. 
A 10\,Gyr, [Fe/H] = $-2.2$ MIST isochrone \citep{d+16, cdc+16, pbd+11, pcb+13, pms+15, psb+18} is over-plotted at the distance modulus of Grus I \citep[$m - M$ = 20.48 mag;][]{cpm+21} for reference.
Top right: Spatial distribution of stars observed in this study. 
The dashed ellipse denotes the Grus I half-light radius presented in \citet{cpm+21}.
Bottom left: Histogram of velocities of the stars in our sample. 
The Grus I members clearly cluster between $\sim-150$\,km\,s$^{-1}$ and $\sim-135$\,km\,s$^{-1}$. 
Note that one star (DES\,J225643.20$-$501130.0) has a velocity consistent with membership but is listed as a non-member due to its high metallicity (see paragraph 2 in Section~\ref{sec:membership}).
Bottom right: \textit{Gaia} EDR3 \citep{gaia+16, gaia+21} proper motions of stars in our sample. 
The members and candidate members of Grus I cluster at a systemic proper motion of $\mu_\alpha\cos\delta = 0.07 \pm 0.05$\,mas\,yr$^{-1}$, $\mu_\delta = 0.25 \pm 0.07$\,mas\,yr$^{-1}$.}
\label{fig:cmd}
\end{figure*}

\subsection{Derivation of radial velocities}
\label{sec:rv}

We derived radial velocities following the methods presented in \citet{sld+17} and \cite{lsd+17}, which we briefly outline here. 
We performed a $\chi^2$ minimization between our observed spectra and a template IMACS spectrum of HD122563 from 8450\,{\AA} to 8680\,{\AA} to measure radial velocities from the calcium triplet features.
The template spectrum of HD 122563 was collected using the same 0\farcs7 slit size and 1200\,$\ell$\,mm$^{-1}$ grating as our Grus~I observations.
We assumed a velocity of $-26.51$\,km\,s$^{-1}$ for HD122563 \citep{cmf+12}. 
We derived heliocentric velocity corrections using the \texttt{astropy} package \citep{astropy, astropy2}.
Random uncertainties on the radial velocity measurements were derived through Monte Carlo re-sampling:
we added noise to each spectrum based on its signal-to-noise value, re-measured the radial velocity, and repeated this process 500 times.
We took the standard deviation of the resulting velocity distribution, after clipping $5\sigma$ outliers, to be the random velocity uncertainty for each star.

We applied a correction to our radial velocities based on the wavelength of the telluric A-band absorption feature at $\sim7600$\,{\AA} to account for slit mis-centering effects. 
We derived this correction by repeating the same steps as for the radial velocity measurement, but instead performing the $\chi^2$ minimization over the wavelength range 7550\,{\AA} to 7700\,{\AA} with respect to a template spectrum of the hot, rapidly rotating star HR4781 \citep{sld+17}. 
These corrections were typically $\lesssim 5$\,km\,s$^{-1}$ and showed a clear dependence on the location of the slit perpendicular to the dispersion axis of the CCD mosaic.
We thereby modeled these corrections by fitting a line to the telluric correction as a function of location along this axis of the CCD mosaic, using only measurements from spectra with S/N $> 5$ to ensure a high-quality sample.
We then used this linear model to calculate the telluric correction for each star, which had the additional benefit of providing a robust correction for stars that had low S/N or no wavelength coverage of the A-band region.

We derived a systematic velocity uncertainty of 1.1\,km\,s$^{-1}$ on our velocity measurements, based on repeat observations of stars following the methods presented in e.g., \citet{sg+07}. 
Specifically, we divided our raw data into two subsets, independently reduced each subset, and derived velocities from the 1D spectra following the above techniques.
Then, we found that a systematic velocity uncertainty of 1.1\,km\,s$^{-1}$ needed to be added in quadrature to the random velocity uncertainties for consistency among the velocity measurement of the same stars \citep[e.g., as in][]{sg+07, sld+17, lsd+17}.
We computed final velocity uncertainties by adding in quadrature the random velocities uncertainties and the systematic velocity uncertainty. 
If applicable (i.e., if stars showed no evidence for binarity), velocity measurements were combined across multiple runs by taking a weighted average, where the weights were equal to the inverse squared uncertainty of the velocity measurement.

Our radial velocity measurements are presented in Table~\ref{tab:gru1spec} and in the lower left panel of Figure~\ref{fig:cmd}. 
We find 13 stars with radial velocities roughly consistent with membership to Grus~I (between $-$150\,km\,s$^{-1}$ and $-$130\,km\,s$^{-1}$).
Of these, nine have metallicity measurements; eight of those have low metallicities ([Fe/H] $< -2.0$) consistent with UFD membership, and one has a higher metallicity  ([Fe/H] $> -1.0$) that is inconsistent with UFD membership (see Figure~\ref{fig:fehvsrv}, and further discussion in Section~\ref{sec:membership}).

\startlongtable
\tabletypesize{\scriptsize}
\begin{deluxetable*}{c c c c c c c r c c}
\tablecaption{Velocity measurements for all stars. \label{tab:gru1spec}}
\tablehead{ID & MJD\tablenotemark{{\scriptsize a}} & RA & DEC & $g$\tablenotemark{{\scriptsize b}} & $r$\tablenotemark{{\scriptsize b}} & S/N & \multicolumn{1}{c}{$v$} &  MEM \\
 &  & (deg) & (deg) & (mag) & (mag) &  & \multicolumn{1}{c}{(\kms)} &   }
\startdata
DES\,J225619.85$-$500757.2 & 58762.03 & 344.08273 & $-$50.13258 & 17.4 & 16.95 & 64.25 & $-4.05 \pm 1.11$ & NM \\
DES\,J225650.20$-$500814.2 & 58762.03 & 344.20917 & $-$50.13729 & 21.06 & 20.57 & 4.85 & $-24.18 \pm 2.57$ & NM \\
DES\,J225625.52$-$500828.1 & 58762.03 & 344.10635 & $-$50.14116 & 21.9 & 20.5 & 10.77 & $-5.37 \pm 1.65$ & NM \\
DES\,J225613.76$-$500835.1 & 58762.03 & 344.05735 & $-$50.14309 & 19.57 & 18.7 & 23.03 & $140.71 \pm 1.26$ & NM \\
DES\,J225713.21$-$500834.6 & 58762.03 & 344.30507 & $-$50.14297 & 20.54 & 19.81 & 10.85 & $51.89 \pm 1.71$ & NM \\
DES\,J225625.24$-$500842.0 & 58762.03 & 344.10519 & $-$50.14501 & 20.35 & 18.87 & 44.23 & $46.26 \pm 1.14$ & NM \\
DES\,J225631.22$-$500841.9 & 58762.03 & 344.13012 & $-$50.14498 & 20.04 & 19.13 & 20.04 & $18.57 \pm 1.8$ & NM \\
DES\,J225710.06$-$500856.0 & 58762.03 & 344.29192 & $-$50.1489 & 19.08 & 18.4 & 27.45 & $0.53 \pm 1.19$ & NM \\
DES\,J225658.06$-$501357.9 & 57229.33 & 344.24192 & $-$50.23276 & 18.55 & 17.49 & 44.95 & $-140.88 \pm 1.11$ & M \\
\nodata & 58762.03 & 344.24192 & $-$50.23276 & 18.55 & 17.49 & 53.25 & $-141.59 \pm 1.11$ & M \\
\nodata & 59471.02 & 344.24192 & $-$50.23276 & 18.55 & 17.49 & 24.79 & $-143.32 \pm 1.13$ & M \\
DES\,J225643.89$-$500903.0 & 58762.03 & 344.18288 & $-$50.15085 & 21.43 & 20.93 & 3.5 & $-144.37 \pm 7.39$ & CM \\
DES\,J225640.97$-$500913.4 & 58762.03 & 344.17074 & $-$50.15374 & 18.97 & 18.63 & 21.8 & $188.02 \pm 1.34$ & NM \\
DES\,J225703.50$-$500942.6 & 58762.03 & 344.26461 & $-$50.16184 & 22.44 & 20.91 & 12.77 & $21.73 \pm 1.92$ & NM \\
DES\,J225720.93$-$500951.2 & 58762.03 & 344.33724 & $-$50.16423 & 21.2 & 19.64 & 30.71 & $33.11 \pm 1.24$ & NM \\
DES\,J225655.34$-$500947.0 & 58762.03 & 344.23059 & $-$50.16308 & 22.01 & 20.5 & 17.49 & $11.87 \pm 1.45$ & NM \\
DES\,J225657.50$-$501013.9 & 58762.03 & 344.23959 & $-$50.17053 & 20.41 & 19.06 & 38.09 & $62.0 \pm 1.15$ & NM \\
DES\,J225710.72$-$501008.0 & 58762.03 & 344.29467 & $-$50.16889 & 22.37 & 21.17 & 3.27 & $60.63 \pm 7.23$ & NM \\
DES\,J225711.61$-$501018.5 & 58762.03 & 344.29838 & $-$50.17182 & 18.99 & 19.12 & 7.75 & $210.38 \pm 6.86$ & NM \\
DES\,J225703.66$-$501016.2 & 58762.03 & 344.26527 & $-$50.17119 & 19.84 & 19.52 & 9.96 & $84.9 \pm 1.79$ & NM \\
DES\,J225632.36$-$501025.4 & 58762.03 & 344.13486 & $-$50.17372 & 20.0 & 19.4 & 10.85 & $57.61 \pm 1.84$ & NM \\
DES\,J225656.51$-$501044.0 & 58762.03 & 344.23547 & $-$50.1789 & 17.55 & 17.2 & 55.85 & $8.95 \pm 1.12$ & NM \\
DES\,J225640.78$-$501051.4 & 58762.03 & 344.16992 & $-$50.18096 & 20.43 & 19.79 & 10.35 & $-142.49 \pm 1.46$ & M \\
\nodata & 59471.02 & 344.16992 & $-$50.18096 & 20.43 & 19.79 & 5.53 & $-140.68 \pm 2.11$ & M \\
DES\,J225717.17$-$501105.1 & 58762.03 & 344.32157 & $-$50.18477 & 19.94 & 19.22 & 12.78 & $79.9 \pm 1.91$ & NM \\
DES\,J225646.98$-$501226.9 & 58762.03 & 344.19576 & $-$50.2075 & 21.96 & 20.78 & 4.93 & $12.44 \pm 3.69$ & NM \\
DES\,J225657.98$-$501252.7 & 58762.03 & 344.24162 & $-$50.21466 & 19.18 & 18.96 & 15.84 & $279.45 \pm 1.6$ & NM \\
DES\,J225629.92$-$500433.3 & 57229.33 & 344.12467 & $-$50.07593 & 19.14 & 18.29 & 33.67 & $-147.39 \pm 1.15$ & M \\
\nodata & 58762.03 & 344.12467 & $-$50.07593 & 19.14 & 18.29 & 36.66 & $-146.38 \pm 1.14$ & M \\
\nodata & 59471.02 & 344.12467 & $-$50.07593 & 19.14 & 18.29 & 15.4 & $-144.72 \pm 1.26$ & M \\
DES\,J225642.47$-$500334.1 & 57229.33 & 344.17699 & $-$50.05949 & 21.58 & 21.06 & 2.02 & $21.47 \pm 10.46$ & NM \\
DES\,J225639.47$-$500401.0 & 57229.33 & 344.16449 & $-$50.06697 & 18.82 & 17.91 & 35.41 & $13.55 \pm 1.16$ & NM \\
DES\,J225619.67$-$500913.1 & 57229.33 & 344.08198 & $-$50.15364 & 20.9 & 20.34 & 6.93 & $-146.35 \pm 2.39$ & M \\
\nodata & 58762.03 & 344.08198 & $-$50.15364 & 20.9 & 20.34 & 6.34 & $-139.67 \pm 2.24$ & M \\
DES\,J225643.20$-$501130.0 & 57229.33 & 344.18001 & $-$50.19168 & 21.13 & 20.48 & 3.69 & $-138.53 \pm 6.06$ & NM \\
\nodata & 58762.03 & 344.18001 & $-$50.19168 & 21.13 & 20.48 & 5.26 & $-138.77 \pm 2.31$ & NM \\
DES\,J225643.79$-$501332.6 & 57229.33 & 344.18246 & $-$50.22574 & 21.51 & 20.95 & 2.57 & $-139.07 \pm 6.87$ & CM \\
DES\,J225637.05$-$501024.8 & 57229.33 & 344.15438 & $-$50.17357 & 20.46 & 19.81 & 9.29 & $-135.25 \pm 2.13$ & M \\
\nodata & 58762.03 & 344.15438 & $-$50.17357 & 20.46 & 19.81 & 9.81 & $-143.08 \pm 1.58$ & M \\
DES\,J225625.69$-$501414.2 & 57229.33 & 344.10707 & $-$50.23729 & 20.95 & 20.42 & 4.35 & $-144.56 \pm 4.0$ & CM \\
DES\,J225653.36$-$500924.3 & 57229.33 & 344.22233 & $-$50.15675 & 19.9 & 19.29 & 13.42 & $-53.12 \pm 1.49$ & NM \\
DES\,J225657.61$-$500938.5 & 57229.33 & 344.24005 & $-$50.1607 & 19.21 & 18.43 & 23.37 & $54.94 \pm 1.29$ & NM \\
\nodata & 58762.03 & 344.24005 & $-$50.1607 & 19.21 & 18.43 & 29.87 & $54.98 \pm 1.2$ & NM \\
DES\,J225658.78$-$500832.5 & 57229.33 & 344.24495 & $-$50.14237 & 20.72 & 20.09 & 5.15 & $166.9 \pm 2.45$ & NM \\
DES\,J225704.98$-$501229.6 & 58762.03 & 344.27078 & $-$50.20824 & 19.88 & 19.26 & 14.21 & $6.57 \pm 1.73$ & NM \\
DES\,J225709.08$-$501214.6 & 57229.33 & 344.28784 & $-$50.20406 & 18.25 & 17.27 & 44.81 & $8.47 \pm 1.14$ & NM \\
DES\,J225709.39$-$500956.8 & 58762.03 & 344.28915 & $-$50.1658 & 21.3 & 20.67 & 4.12 & $118.66 \pm 2.35$ & NM \\
DES\,J225722.15$-$501150.7 & 57229.33 & 344.34232 & $-$50.19744 & 19.73 & 18.91 & 10.27 & $-29.24 \pm 1.7$ & NM \\
DES\,J225642.95$-$501741.3 & 57229.33 & 344.17897 & $-$50.29482 & 21.18 & 20.62 & 3.16 & $-148.87 \pm 4.97$ & CM \\
DES\,J225643.29$-$500607.3 & 57229.33 & 344.18038 & $-$50.10203 & 19.74 & 19.07 & 17.25 & $-146.79 \pm 1.25$ & M \\
\nodata & 58762.03 & 344.18038 & $-$50.10203 & 19.74 & 19.07 & 20.64 & $-146.41 \pm 1.24$ & M \\
DES\,J225611.70$-$500304.8 & 58762.03 & 344.04878 & $-$50.05136 & 19.86 & 19.31 & 14.82 & $-85.61 \pm 1.35$ & NM \\
DES\,J225617.78$-$500309.3 & 58762.03 & 344.0741 & $-$50.05259 & 20.42 & 19.21 & 20.75 & $116.71 \pm 1.38$ & NM \\
DES\,J225602.89$-$500353.8 & 58762.03 & 344.01204 & $-$50.06496 & 20.05 & 18.61 & 57.85 & $-27.75 \pm 1.13$ & NM \\
DES\,J225618.15$-$500405.3 & 58762.03 & 344.07565 & $-$50.06815 & 18.94 & 18.66 & 21.47 & $-31.7 \pm 1.3$ & NM \\
DES\,J225615.03$-$500412.8 & 58762.03 & 344.06263 & $-$50.07025 & 21.7 & 20.33 & 12.92 & $137.47 \pm 1.56$ & NM \\
DES\,J225602.84$-$500414.7 & 58762.03 & 344.01186 & $-$50.07076 & 21.48 & 21.06 & 2.6 & $144.89 \pm 5.46$ & NM \\
DES\,J225619.00$-$500427.2 & 58762.03 & 344.0792 & $-$50.07423 & 18.52 & 18.02 & 34.99 & $-22.89 \pm 1.15$ & NM \\
DES\,J225633.63$-$500426.1 & 58762.03 & 344.14012 & $-$50.07393 & 21.98 & 20.55 & 12.11 & $-6.86 \pm 1.67$ & NM \\
DES\,J225626.96$-$500445.3 & 58762.03 & 344.11236 & $-$50.07927 & 21.77 & 20.68 & 5.73 & $-52.27 \pm 4.14$ & NM \\
DES\,J225615.78$-$500452.8 & 58762.03 & 344.06578 & $-$50.08134 & 21.23 & 20.9 & 2.82 & $-48.25 \pm 4.49$ & NM \\
DES\,J225601.68$-$500459.2 & 58762.03 & 344.00702 & $-$50.08312 & 22.33 & 20.69 & 15.95 & $-5.16 \pm 1.46$ & NM \\
DES\,J225603.76$-$500524.5 & 58762.03 & 344.0157 & $-$50.09014 & 20.94 & 20.31 & 5.53 & $-141.21 \pm 1.67$ & M \\
DES\,J225624.48$-$500540.8 & 58762.03 & 344.10202 & $-$50.09468 & 22.19 & 20.61 & 17.33 & $8.0 \pm 1.47$ & NM \\
DES\,J225638.97$-$500542.4 & 58762.03 & 344.1624 & $-$50.09511 & 20.02 & 19.43 & 14.25 & $20.45 \pm 1.82$ & NM \\
DES\,J225644.70$-$500559.7 & 58762.03 & 344.18625 & $-$50.09993 & 21.95 & 20.63 & 11.68 & $35.89 \pm 1.96$ & NM \\
DES\,J225608.69$-$500603.8 & 58762.03 & 344.03622 & $-$50.10107 & 19.66 & 19.19 & 14.54 & $50.85 \pm 1.65$ & NM \\
DES\,J225628.96$-$500601.9 & 58762.03 & 344.12068 & $-$50.10054 & 20.97 & 20.0 & 9.75 & $89.51 \pm 1.83$ & NM \\
DES\,J225613.54$-$500603.5 & 58762.03 & 344.05645 & $-$50.10099 & 21.05 & 19.92 & 14.86 & $-19.48 \pm 1.48$ & NM \\
DES\,J225613.72$-$500616.4 & 58762.03 & 344.05718 & $-$50.10458 & 16.59 & 16.75 & 52.03 & $156.24 \pm 1.25$ & NM \\
DES\,J225649.23$-$501031.4 & 58762.03 & 344.20513 & $-$50.17539 & 20.79 & 20.22 & 6.91 & $-142.47 \pm 1.83$ & M \\
DES\,J225603.42$-$500617.4 & 58762.03 & 344.01426 & $-$50.10483 & 19.07 & 18.68 & 19.51 & $58.47 \pm 1.25$ & NM \\
DES\,J225601.57$-$500623.0 & 58762.03 & 344.00655 & $-$50.1064 & 21.35 & 19.93 & 18.73 & $76.15 \pm 1.32$ & NM \\
DES\,J225617.16$-$500621.4 & 58762.03 & 344.0715 & $-$50.10596 & 22.54 & 21.44 & 2.8 & $6.05 \pm 9.32$ & NM \\
DES\,J225555.38$-$500642.3 & 58762.03 & 343.98078 & $-$50.11175 & 18.55 & 17.68 & 44.28 & $-70.77 \pm 1.14$ & NM \\
DES\,J225626.51$-$500651.4 & 58762.03 & 344.11047 & $-$50.1143 & 19.98 & 18.53 & 51.53 & $11.07 \pm 1.12$ & NM \\
DES\,J225616.50$-$500651.2 & 58762.03 & 344.06877 & $-$50.11425 & 21.05 & 20.64 & 3.98 & $142.27 \pm 3.55$ & NM \\
DES\,J225653.50$-$500716.3 & 58762.03 & 344.22292 & $-$50.12122 & 21.28 & 20.22 & 8.67 & $27.15 \pm 2.49$ & NM \\
DES\,J225657.01$-$500740.7 & 58762.03 & 344.23755 & $-$50.12799 & 18.09 & 17.65 & 43.95 & $-49.15 \pm 1.17$ & NM \\
DES\,J225644.73$-$500749.0 & 58762.03 & 344.1864 & $-$50.1303 & 21.99 & 20.7 & 7.64 & $98.16 \pm 2.55$ & NM
\enddata
\tablenotetext{a}{the midpoint MJD of observation.}
\tablenotetext{b}{Quoted magnitudes represent the weighted-average dereddened PSF magnitude derived from the DES DR2 catalog \citep{fdh+15, mgm+18, aaa+21}.}
\tablenotetext{c}{NM indicates non-members, CM indicates candidate members (stars with consistent radial velocity, but no derived metallicity), M indicates members (see Section~\ref{sec:membership}).}
\end{deluxetable*}

\subsection{Derivation of metallicities}
\label{sec:feh}

We derived metallicities by using the well-established relationship between the equivalent widths of the calcium triplet absorption lines, the absolute $V$ magnitude, and the stellar metallicity \citep{cpg+13}. 
We measured equivalent widths of each calcium triplet line by fitting them with a Gaussian plus Lorentzian profile, following \citet{sld+17} and \citet{lsd+17}.
The apparent $V$ magnitude of each star was derived using Equation~5 in \citet{bdb+15} and was converted to an absolute magnitude using a Grus I distance modulus of $m-M=20.48$ \citep{cpm+21}.

The random uncertainties in the metallicities were derived following the procedure in \citet{sle+20}, in which the statistical uncertainty in the equivalent width from the Gaussian + Lorentzian fit to each line was propagated to derive a metallicity uncertainty. 
An additional systematic uncertainty of 0.32{\,\AA} was added in quadrature to the equivalent width uncertainties \citep[e.g.,][]{sld+17}.
The systematic uncertainty in the final metallicities was assumed to be 0.17\,dex, following the stated uncertainty of the metallicity calibration in \citet{cpg+13}. 
We took the final metallicity uncertainty of each star as the quadrature sum of its random uncertainty and the systematic uncertainty.
For stars with spectra over multiple epochs, we combined metallicity measurements from the individual spectra by taking the weighted average of the metallicities, where the weights were the inverse square of the metallicity uncertainties.

\begin{figure}[!htbp]
\centering
\includegraphics[width =\columnwidth]{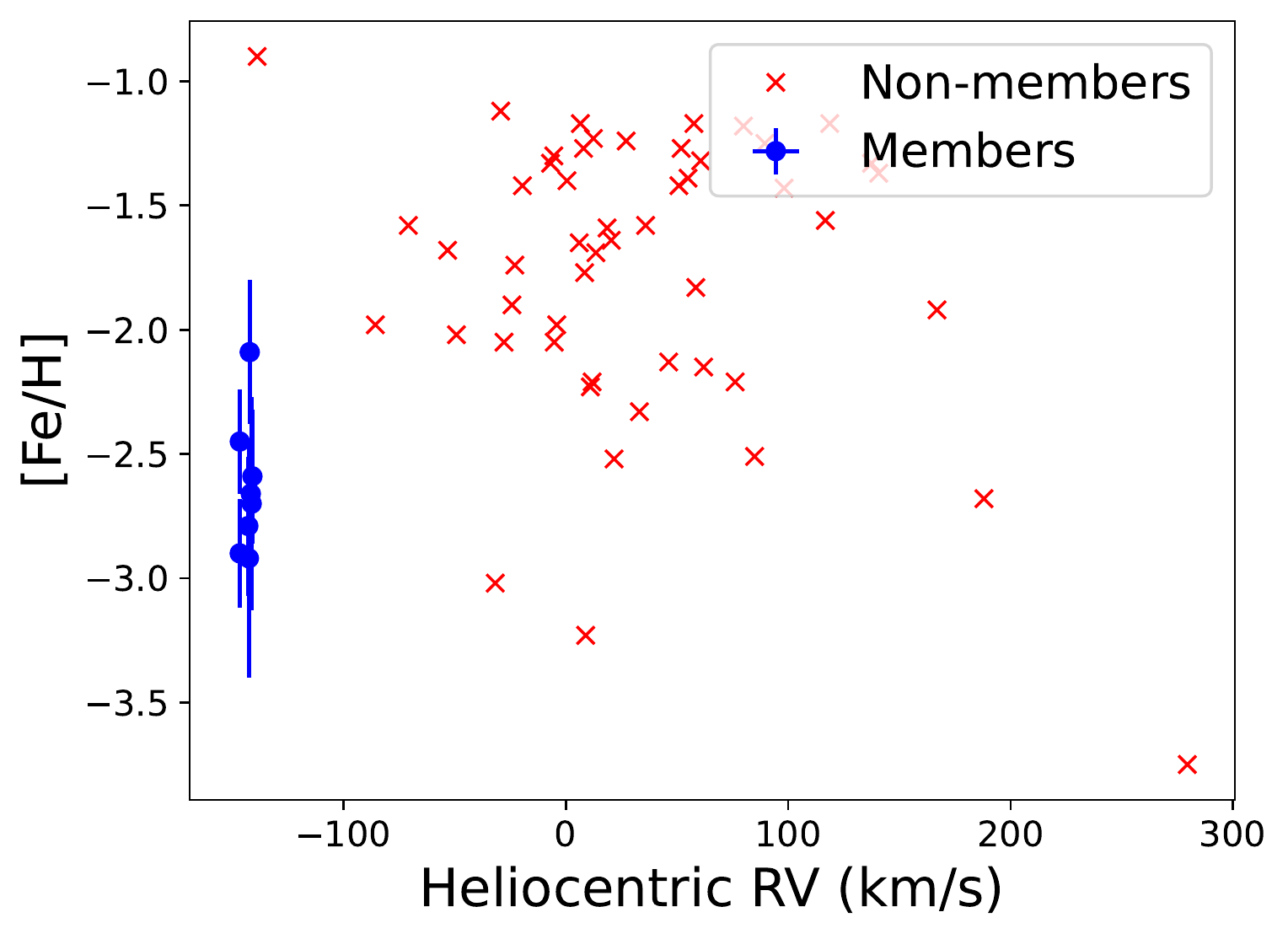}
\caption{[Fe/H] vs. Radial velocities for stars in this study. 
The members of Grus I clearly cluster at low metallicities ([Fe/H] $< -2.0$) and at a systematic velocity of $\sim-143$\,km\,s$^{-1}$. 
Note that one star has a velocity consistent with membership but a high metallicity ([Fe/H] $-0.9 \pm 0.38$\,dex), and is therefore listed as a non-member. 
There are additional candidate members in our sample with velocities consistent with membership (see Section~\ref{sec:membership}), but are not shown in this plot as the S/N of their spectra is too low for a metallicity determination.}
\label{fig:fehvsrv}
\end{figure}

We note that the calibration we employ is only valid for red giant branch stars that are at the distance of Grus~I, since the absolute $V$ magnitudes inputted into the calibration are computed assuming the distance modulus of Grus~I. 
We see an artifact of this assumption in Figure~\ref{fig:fehvsrv}, where a number of non-members of Grus I (e.g., stars with radial velocities inconsistent with membership) have spuriously low metallicities. 
Accordingly, we only report metallicities in Table~\ref{tab:gru1feh} for stars that are confirmed members of Grus I (defined as radial velocities consistent with membership and a low metallicity; see Section~\ref{sec:membership}). 

\subsection{Identifying members of Grus~I}
\label{sec:membership}

We identify Grus I members based on their clustered radial velocities and the fact that UFD stars have low metallicities ([Fe/H] $\lesssim -1.5$; \citealt{s+19}).
\citet{wmo+16} find that Grus I has a systematic radial velocity of $-140.5^{+2.4}_{-1.6}$\,km\,s$^{-1}$.
We find a significant over-density of stars at a similar radial velocity, as shown in the  bottom left panel of Figure~\ref{fig:cmd}. 
We select all stars with radial velocities between $-150$\,km\,s$^{-1}$ and $-130$\,km\,s$^{-1}$ as an initial sample of 13 possible members. 
There are no stars with velocities just beyond the threshold of these limits (i.e., no other stars have velocities $< -100$\,km\,s$^{-1}$), so it is unlikely that we have excluded any possible members with this velocity cut.
We note that all of the stars with radial velocities that are consistent with Grus~I membership also have \textit{Gaia} EDR3 proper motions \citep{gaia+16, gaia+21}, when available, consistent with the systematic proper motion of Grus~I ($\mu_\alpha \cos \delta = 0.07 \pm 0.05$\,mas\,yr$^{-1}$, $\mu_\delta = -0.29^{+0.06}_{-0.07}$\,mas\,yr$^{-1}$; \citealt{mv+20}).
This consistency is illustrated in the bottom right panel of Figure~\ref{fig:cmd}.

Of the 13 stars with velocities consistent with Grus~I membership, 9 have spectra with sufficient signal-to-noise (S/N $> 5$) to derive metallicities from the calcium triplet absorption features. 
We find that 8 of these 9 stars have very low metallicities ([Fe/H] $< -2.0$), consistent with UFD membership (see Figures 5 and 6 in \citealt{s+19}). 
Stars at such low metallicities are unlikely to be foreground Milky Way stars \citep[e.g.][]{ysm+20, cmf+21}.
One star (DES\,J225643.20$-$501130.0) has a velocity and proper motion consistent with membership, but a high metallicity of [Fe/H] = $-0.9 \pm 0.38$ which is above the typical metallicity range of UFD stars (no known UFD star has [Fe/H] $> -1.0$). 
Based on its high metallicity, we therefore identify this star as a non-member and exclude it from further analysis, but we note for completeness that none of the primary conclusions of our paper (e.g., orbital properties, Grus I being dark-matter-dominated) would change if we were to include this star.
As noted in Section~\ref{sec:feh}, the metallicity calibration that we employ is only valid for stars at the distance of Grus~I. 
Accordingly, the metallicities of stars that are likely not members of Grus~I should be disregarded.

We identify the 8 stars with low metallicities ([Fe/H] $< -2.0$) and velocities consistent with Grus~I membership ($-150$\,km\,s$^{-1}$ to $-130$\,km\,s$^{-1}$) as confirmed members (M in Table~\ref{tab:gru1spec}).
We report the 4 stars with velocities consistent with membership, but no metallicity measurements as candidate members (CM in Table~\ref{tab:gru1spec}).
All other stars are identified as non-members.
The analysis in Section~\ref{sec:discussion} is performed using the sample of confirmed members.

\subsection{Combining our sample with existing literature measurements}
\label{sec:combining}

There are two published studies of Grus~I that report velocity and metallicity values of individual member stars \citep{wmo+16, jsf+19}.
We opt to incorporate some of those measurements in our analyses of Grus I to increase our sensitivity to the velocity dispersion of the system and to aid in detecting binary stars.
In this subsection, we outline how our velocities and metallicities compare to values presented in those studies, and then discuss whether/how we incorporate their measurements into our study.

\subsubsection{Comparison to \citet{wmo+16}}
\label{sec:m2fs}

\citet{wmo+16} used the M2FS multi-fiber instrument on the Magellan/Clay telescope to identify 7 likely members of Grus I ($p_{\text{mem}} > 0.5$ in their Table 1). 
We re-observed all of those stars (Gru1-003, Grus-004, Grus1-007, Gru1-023, Gru1-032, Gru1-035, Gru1-038) in this study. 
We identify two stars in \citealt{wmo+16} (Gru1-022 and Gru1-054) that they do not classify as likely members to be members with our IMACS data (DES\,J225619.67$-$500913.1 and DES\,J225603.76$-$500524.5, respectively, in Table~\ref{tab:gru1spec}).
We find that both stars have IMACS velocities consistent with membership (between $-143$\,km\,s$^{-1}$ and $-141$\,km\,s$^{-1}$) and low metallicities ([Fe/H] $< -2.5$).
They likely missed being classified as members in \citet{wmo+16} due to their large velocity uncertainties (19.2\,km\,s$^{-1}$ and 87.9\,km\,s$^{-1}$, respectively) in that study.
We note that we identify one likely member in \citet{wmo+16}, Gru1-007 (DES\,J225643.20$-$501130.0), as a non-member due to its high IMACS metallicity of [Fe/H] = $-0.9 \pm 0.38$ (see second paragraph in Section~\ref{sec:membership}). 
The star also has a relatively high metallicity of [Fe/H] $= -1.19 \pm0.42$ in \citet{wmo+16} and is marginally redder than the rest of the Grus~I members (as denoted by the larger red cross in the top left panel of Figure~\ref{fig:cmd}).

The most notable discrepancy between our study and \citet{wmo+16} is in the metallicity measurements.
Six stars we classify as confirmed members have M2FS metallicities in their study (Gru1-003, Gru1-004, Gru1-022, Gru1-032, Gru1-038, Gru1-054). 
Our metallicities are, on average, 0.45$\pm0.14$\,dex lower than their M2FS metallicities. 
Notably, Gru1-003 (DES\,J225637.05$-$501024.8), Gru1-004 (DES\,J225640.78$-$501051.4), Gru1-022 (DES\,J225619.67$-$500913.1), and Gru1-054 (DES\,J225603.76$-$500524.5) have discrepancies in their metallicities of over 0.5\,dex; although, Gru1-003, Gru1-022, and Gru1-054 also have highly uncertain M2FS metallicities ($\sigma \geq 0.4$\,dex) in \citet{wmo+16}. 
The overall offset between our metallicities and those in \citet{wmo+16} is largely explained by the systematic metallicity offset of $+0.32$\,dex that \citet{wmo+16} added to their metallicities to account for discrepancies with solar values.
This offset has been discussed in \citet{cfj+18} and \citet{jsf+19} as a cause of discrepancy between M2FS metallicities and those derived from high-resolution spectroscopy.
Accordingly, we adopt our IMACS calcium triplet-based metallicities in all subsequent analyses. 

We find a small, but statistically significant, average velocity offset of $v_{\text{IMACS}} - v_{\text{M2FS}} = -2.6 \pm 0.8$\,km\,s$^{-1}$ between the IMACS velocities of the Grus~I candidate members and the M2FS velocities presented in \citet{wmo+16}.
After accounting for this offset, there are no $>2\sigma$ velocity outliers between our studies, reflecting agreement between the M2FS and IMACS velocities.
We note that attempting to model this offset as part of our dynamical modeling (see Section~\ref{sec:dynamics}) returns an offset of $v_{\text{IMACS}} - v_{\text{M2FS}} = -2.8^{+1.0}_{-0.9}$\,km\,s$^{-1}$, consistent with this more direct estimate of the offset.

We opt to combine our velocities with the M2FS velocities in \citet{wmo+16} to increase the velocity precision of our sample. 
To account for possible systematic effects when combining measurements from different spectrographs, we implement the likelihood function presented in \citet{mpm+19} in our dynamical analysis in Section~\ref{sec:dynamics}.
This likelihood function simultaneously fits for a velocity offset between samples from different spectrographs when deriving dynamical parameters.
Additionally, we add a systematic velocity uncertainty of 0.9\,km\,s$^{-1}$ in quadrature to the velocity uncertainties presented in \citet{wmo+16}, following the analysis of M2FS velocity uncertainties in \citet{sdl+15}. 
We note that the net effect of both of these corrections is to decrease the significance of any detected velocity dispersion, making these conservative choices with respect to the conclusions of this paper.
We present results both with and without these kinematic adjustments in Section~\ref{sec:dynamics}, but the relevant numbers in Table~\ref{tab:gru1table} and elsewhere in the paper reflect the steps that are described in this paragraph.

\subsubsection{Comparison to \citet{jsf+19}}
\label{sec:mike}

\citet{jsf+19} used the MIKE spectrograph on Magellan/Clay to obtain high-resolution spectra of Gru1-032 (DES\,J225658.06$-$501357.9) and Gru1-038 (DES\,J225629.92$-$500433.3) to derive their detailed abundances. 
We find excellent agreement (within 0.1\,dex) between our metallicities and their metallicities.
We opt to use our calcium triplet-based metallicities for these stars in the analysis in this paper to ensure uniformity in how metallicities are derived across our sample.
Moreover, given the agreement between the calcium triplet metallicities and the MIKE metallicities, opting for one set over the other does not change any results. 

We re-measure the velocities of Gru1-032 (DES\,J225658.06$-$501357.9) and Gru1-038 (DES\,J225629.92$-$500433.3) from the MIKE spectra presented in \citet{jsf+19}, following the steps outlined in \citet{cfj+22}.
We derive a velocity of $v_{\text{MIKE}} = -139.6 \pm 1.2$\,km\,s$^{-1}$ for Gru1-032, and $v_{\text{MIKE}} = -143.3 \pm 1.2$\,km\,s$^{-1}$ for Gru1-038. 
These are within $4$\,km\,s$^{-1}$ of the IMACS and M2FS velocities.
We incorporate these velocity measurements in our binarity analysis in Section~\ref{sec:binaries}.

\subsection{Identifying binary stars in Grus~I}
\label{sec:binaries}

We searched for evidence of binarity in our Grus~I candidate members by combining velocity data from this study, \citet{wmo+16}, and \citet{jsf+19}.
Before performing this analysis, we applied an offset of $-2.6$\,km\,s$^{-1}$ to the velocities in \citealt{wmo+16} (see paragraph 3 in Section~\ref{sec:m2fs}).
We also added a systematic velocity uncertainty of 0.9\,km\,s$^{-1}$ in quadrature to the uncertainties provided in \citealt{wmo+16} (see paragraph 4 in Section~\ref{sec:m2fs}).
The MIKE velocities from \citet{jsf+19} for Gru1-032 (DES\,J225658.06$-$501357.9) and Gru1-038 (DES\,J225629.92$-$500433.3) were taken as the values presented in the second paragraph of Section~\ref{sec:mike}.
We note that \citet{jsf+19} do not report evidence that either Gru1-032 or Gru1-038 are binaries when comparing their MIKE velocities to M2FS velocities of those stars in \citet{wmo+16}.

We tested for binarity by performing a $\chi^2$ test on each star to test the null hypothesis that its velocity is constant over time.
The IMACS velocities used for this test are provided in Table~\ref{tab:gru1spec}, and the M2FS and MIKE velocities were included when available.
We note that DES\,J225649.23$-$501031.4 could not be tested for binarity since it only has a usable radial velocity measurement from one epoch.
The same is effectively true for DES\,J225603.76$-$500524.5/Gru1-054, which only has an IMACS velocity from one epoch and a highly uncertain M2FS velocity.
We find strong evidence ($p = 0.01$) of binarity for Gru1-003 (DES\,J225637.05$-$501024.8), and marginal evidence ($p = 0.04$) for Gru1-022 (DES\,J225619.67$-$500913.1) if one excludes its uncertain M2FS velocity ($\sigma=$19.2\,km\,s$^{-1}$).
To be conservative, we exclude velocities from these stars in our dynamical analysis of Grus I in Section~\ref{sec:dynamics}.

\section{Discussion}
\label{sec:discussion}

\begin{deluxetable}{llr}
\tablecaption{\label{tab:gru1table}Summary of Properties of Grus\,I}
\tablehead{
\colhead{Row} & \colhead{Quantity} & \colhead{Value}
}
\startdata
(1) & RA (J2000)                           & $344.166^{+0.007}_{-0.006}$ \\
(2) & Dec (J2000)                          & $-50.168^{+0.006}_{-0.005}$ \\
(3) & Distance (kpc)                       & $125^{+6}_{-12}$  \\
(4) & $m - M$ (mag)                        & $20.48^{+0.11}_{-0.22}$\\
(5) & $r_{\rm 1/2}$ (arcmin)               & $4.16^{+0.54}_{-0.74}$ \\
(6) & Ellipticity                          & $0.44^{+0.08}_{-0.10}$ \\
(7) & Position angle (degrees)             & $153^{+8}_{-7}$ \\
(8) & $M_{V,0}$                             & $-4.1\pm0.3$ \\
(9) & $r_{\rm 1/2}$ (pc)                    & $151^{+21}_{-31}$  \\ 
\hline
(10)  & N$_{\rm spectroscopic~members}$         & 8\tablenotemark{a} \\
(11)  & $V_{\rm hel}$ (\kms)                  & $-143.5^{+1.2}_{-1.2}$\,km\,s$^{-1}$\\
(12)  & $V_{\rm GSR}$ (\kms)                  & $-189.4^{+1.2}_{-1.2}$\,km\,s$^{-1}$ \\
(13)  & $\sigma$ (\kms)                       & $2.5^{+1.3}_{-0.8}$\,km\,s$^{-1}$\\
(14)  & Mass (M$_{\odot}$)                   & $8^{+12}_{-4} \times 10^5$  \\
(15)  & M/L$_{V}$ (M$_{\odot}$/L$_{\odot}$) & 440$^{+650}_{-250}$ \\
(16)  & Mean [Fe/H]                        & $-2.62 \pm 0.11$ \\
(17)  & $\mua$ (\masyr)                    & $0.07 \pm 0.05$\,mas\,yr$^{-1}$ \\
(18)  & $\mud$ (\masyr)                    & $-0.25 \pm 0.07$\,mas\,yr$^{-1}$ \\
(19)  & Orbital pericenter (kpc)              & 49$^{+27}_{-23}$\,kpc\tablenotemark{b} \\
(20)  & Orbital apocenter (kpc)               & 205$^{+58}_{-24}$\,kpc\tablenotemark{b} \\
(21)  & $\log_{10}{J(0.2\degr)}$ (GeV$^{2}$~cm$^{-5}$) & $16.4_{-0.7}^{+0.8}$ \\
(22)  & $\log_{10}{J(0.5\degr)}$ (GeV$^{2}$~cm$^{-5}$) & $16.5_{-0.7}^{+0.8}$ \\
\enddata
\tablecaption CColumns (1) through (9) are from \citet{cpm+21}; Columns (19) and (20) are from \citet{pel+22}; all other Columns are from this study.
\tablenotetext{a}{Only includes stars with a confirmed low metallicity from IMACS spectroscopy.}
\tablenotetext{b}{Taken from \citealt{pel+22}, which presented Grus~I orbital parameters including the effect of the LMC, but with the systemic radial velocity in \citet{wmo+16}. Our updated systemic velocity should negligibly affect these values; see Section~\ref{sec:orbit} for more discussion.}
\end{deluxetable}


In this Section, we answer four questions about Grus~I: (1) What is its dynamical mass and is it dark-matter-dominated? (2) Does Grus I follow the mass-metallicity and metallicity-luminosity relations for dwarf galaxies? (3) What is the viability of using Grus I for searches for dark matter (DM) interactions? And (4) What is the orbital history of Grus I?
We then conclude by comparing our results to those currently in the literature.
In subsequent analysis, we use the IMACS metallicities and velocities of confirmed Grus~I members (as described in Section~\ref{sec:membership}), supplemented by their M2FS velocity measurements in \citealt{wmo+16} (see Section~\ref{sec:m2fs}).

\begin{figure*}[!htbp]
\centering
\includegraphics[width =\columnwidth]{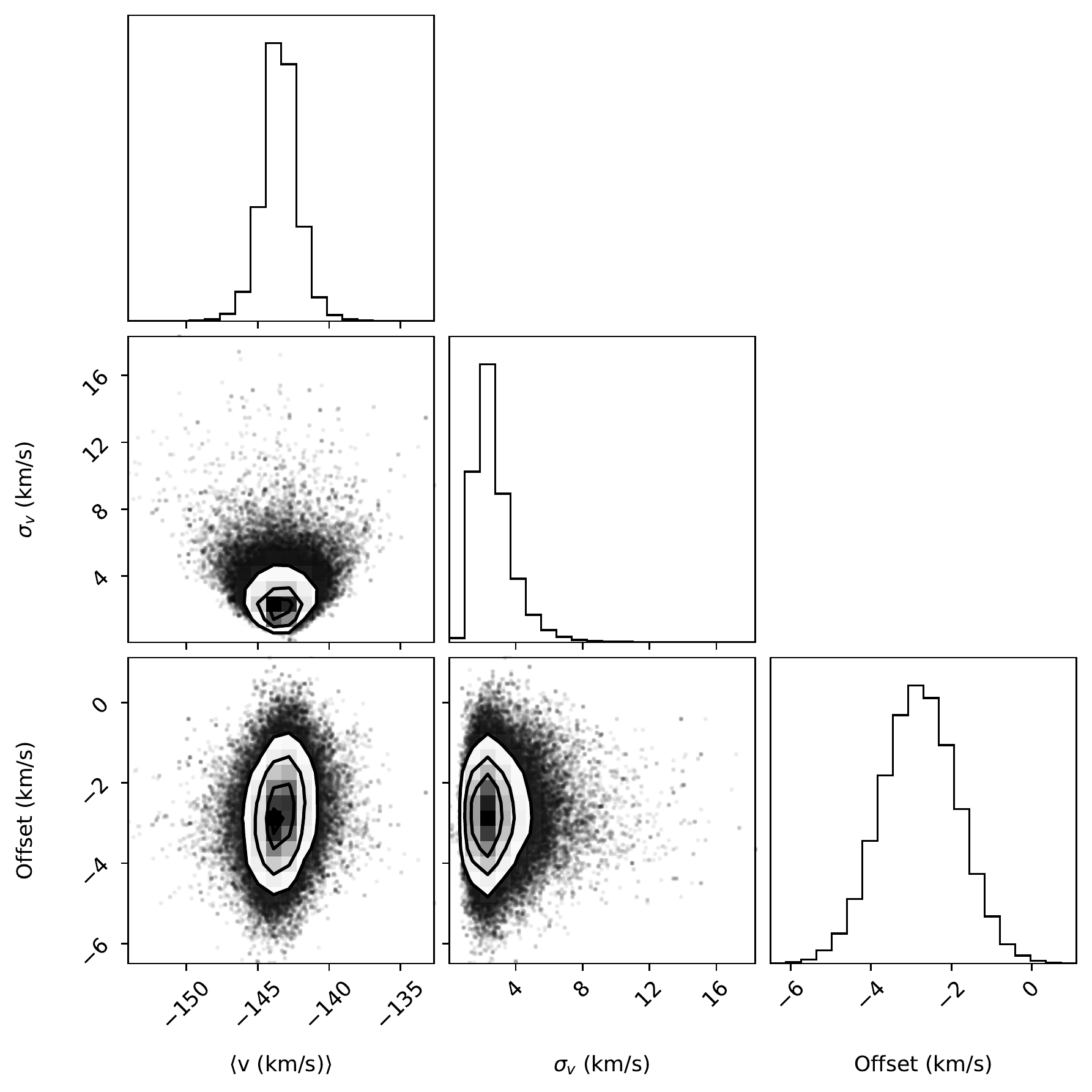}
\includegraphics[width =\columnwidth]{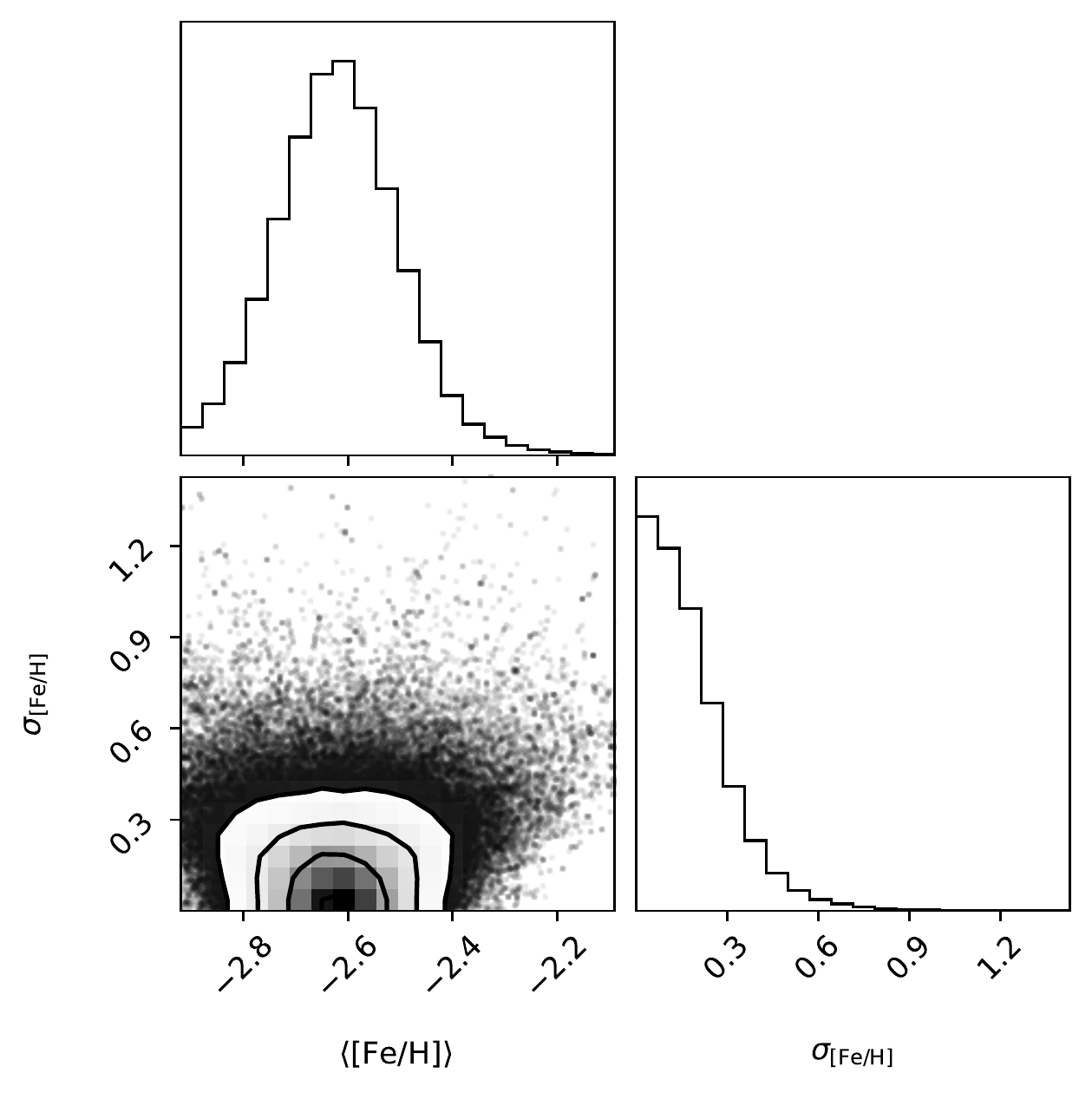}
\caption{Left panels: Corner plot from our Markov Chain Monte Carlo (MCMC) analysis to determine the systemic radial velocity and velocity dispersion of Grus I, while jointly fitting for the velocity offset between the M2FS and IMACS spectrographs (see Section~\ref{sec:dynamics}).
We determine a systematic velocity of $-143.5\pm1.2$\,km\,s$^{-1}$ and a velocity dispersion of $2.5^{+1.3}_{-0.8}$\,km\,s$^{-1}$, confirming Grus I to be dark matter-dominated ($>80$\,M$_\odot$/$L_\odot$ from the 95\% confidence interval; see Section~\ref{sec:dynamics}) as is typical of UFDs. 
Right panels: Corner plot from our MCMC analysis to determine the mean metallicity and metallicity dispersion of Grus I.
We find that Grus I has a low $\langle$[Fe/H]$\rangle$ = $-2.62 \pm 0.11$, also as is typical of UFDs, and place an upper limit on the metallicity dispersion of $\sigma_{\text{[Fe/H]}} < 0.44$.}
\label{fig:cornerplots}
\end{figure*}

\subsection{Dynamical mass of Grus~I}
\label{sec:dynamics}

We derive the dynamical mass within a half-light radius of Grus I using the estimator presented in \citet{wmb+10}:

\begin{equation}
\label{eqn:mass}
M (r_{1/2}) \simeq \left(\frac{\langle\sigma_{\text{los}}^2\rangle}{\text{km}^2\,\text{s}^{-2}}\right)\,\left(\frac{R_e}{\text{pc}} \right)\,M_\odot
\end{equation}
in which $\langle\sigma_{\text{los}}^2\rangle$ is the squared line-of-sight velocity dispersion and $R_{\text{e}}$ is the two-dimensional projected half-light radius.
We adopt $R_{\text{e}} = 151^{+21}_{-31}$\,pc \citep{cpm+21} in all subsequent calculations.

We derive the velocity dispersion of Grus~I using a maximum-likelihood approach on a joint sample of IMACS and M2FS \citep{wmo+16} velocity measurements. 
We restrict our sample to confirmed members (as described in the last paragraph of Section~\ref{sec:membership}) and stars that do not show evidence of binarity (see Section~\ref{sec:binaries}). 
These restrictions result in a sample of six stars for the velocity dispersion derivation. 
Three of these stars (DES\,J225658.06, DES\,J225640.78, and DES\,J225629.92 in Table~\ref{tab:gru1feh}) have precise M2FS velocities (uncertainties $< 2$\,km\,s$^{-1}$) in \citet{wmo+16}.
To self-consistently incorporate these M2FS velocities with our IMACS velocities when deriving the velocity dispersion, we implement the likelihood function presented in equations 2 and 3 of \citet{mpm+19}.
This likelihood simultaneously fits for velocity offsets in velocities from different spectrographs when deriving the systemic velocity and velocity dispersion. 
We implemented this likelihood function in \texttt{emcee} \citep{emcee1, emcee2}, and initialized our sample with 100 walkers with a uniform prior on the velocity offset and systemic velocity, and a Jeffreys Prior on the velocity dispersion.
The resulting corner plot after 2000 steps is shown in the left panels of Figure~\ref{fig:cornerplots}. 

From this MCMC analysis (and as seen in Figure~\ref{fig:cornerplots}), we derive a systemic velocity of $-143.5\pm1.2$\,km\,s$^{-1}$, a velocity dispersion of $2.5^{+1.3}_{-0.8}$\,km\,s$^{-1}$, and a velocity offset between measurements from the M2FS and IMACS spectrographs of $-2.8_{-0.9}^{+1.0}$\,km\,s$^{-1}$. 
The 95\% confidence interval (CI) of the velocity dispersion is 1.2\,km\,s$^{-1}$ to 6.2\,km\,s$^{-1}$, demonstrating that we clearly resolve a velocity dispersion at a $>2\sigma$ level.
This dispersion results in a dynamical mass within a half-light radius of 8$^{+12}_{-4} \times 10^5$\,M$_\odot$ (95\% CI of 1.7$ \times 10^5$\,M$_\odot$ to 5.2$ \times 10^6$\,M$_\odot$).
The corresponding mass-to-light ratio is 440$^{+650}_{-250}$\,M$_\odot$/L$_\odot$ (95\% CI of 80\,M$_\odot$/L$_\odot$ to 3000\,M$_\odot$/L$_\odot$) using the absolute magnitude of $M_V = -4.1 \pm 0.3$ reported in \citet{cpm+21}. 
This establishes that Grus I is a canonical, dark-matter-dominated UFD.

To ensure this conclusion is robust to how we combined the M2FS and IMACS velocities, we repeat the above analysis with several modifications and present the results here. 
When including the metal-rich star at the systemic velocity of Grus I in our analysis (DES\,J225643.20$-$501130.0; see paragraph 2 in Section~\ref{sec:membership}), we derive a systemic velocity of $-143.1^{+1.2}_{-1.0}$\,km\,s$^{-1}$ and a velocity dispersion of $2.5^{+1.3}_{-0.7}$\,km\,s$^{-1}$.
If we do not add the systematic velocity uncertainty of 0.9\,km\,s$^{-1}$ to the M2FS velocities and repeat the above analysis, we derive a systemic velocity of $-143.5^{+1.3}_{-1.2}$\,km\,s$^{-1}$ and a velocity dispersion of $2.7^{+1.3}_{-0.8}$\,km\,s$^{-1}$. 
If we choose to combine the M2FS and IMACS velocities by just manually adding an offset of $-2.6$\,km\,s$^{-1}$ to the M2FS velocities, taking a weighted average with the IMACS velocities, and repeating the analysis, we derive a systematic velocity of $-143.5^{+1.2}_{-1.1}$\,km\,s$^{-1}$ and a velocity dispersion of $2.5^{+1.2}_{-0.8}$\,km\,s$^{-1}$.
If we include the two binary candidates (DES\,J225637.05$-$501024.8 and DES\,J225619.67$-$500913.1) using their mean velocities, we derive a dispersion of $2.3^{+0.9}_{-0.7}$\,km\,s$^{-1}$; similarly, if we also include the four candidate members (see Section~\ref{sec:membership}), we derive a velocity dispersion of $2.1^{+0.7}_{-0.6}$\,km\,s$^{-1}$. 
Including the MIKE velocities of DES\,J225658.06$-$501357.9 and DES\,J225629.92$-$500433.3 still results in a significant dispersion of $2.3^{+1.2}_{-0.7}$\,km\,s$^{-1}$.
The only case that results in the system not being dark-matter-dominated at the 2$\sigma$ level is when only the IMACS velocities of the six confirmed members are used; this sample results in a dispersion of $2.1^{+1.3}_{-0.9}$\,km\,s$^{-1}$ and a 95\% confidence interval on the mass-to-light ratio from 3\,M$_\odot$/L$_\odot$ to 2400\,M$_\odot$/L$_\odot$. 
However, in the latter case, the system again becomes clearly M/L$_V$ dominated when a uniform prior is used instead of Jeffrey's prior when deriving the dispersion. 
Accordingly, in all but one case, none of the Grus~I derived properties are meaningfully sensitive to our choice of how to combine the M2FS and IMACS datasets. 
For completeness, we also note that tests with mock data show that we can recover the correct velocity dispersion of Grus I with a sample of six stars, with similar uncertainties to what we obtained.

\subsection{Metallicity properties of Grus~I}
\label{sec:fehspread}

We derive the mean metallicity and metallicity dispersion of Grus~I using the metallicities of its eight confirmed members (see Table~\ref{tab:gru1feh}).
We implement the exact same MCMC approach and implementation as in Section~\ref{sec:dynamics}, but instead just use the metallicity-related terms in the likelihood function in equation 4 of \citet{wmo+16} assuming no metallicity gradient. 
This likelihood models the metallicity distribution as a Gaussian with a mean metallicity $\mu_{\text{[Fe/H]}}$ and a metallicity dispersion $\sigma_{\text{[Fe/H]}}$.

\begin{figure}[!htbp]
\centering
\includegraphics[width =\columnwidth]{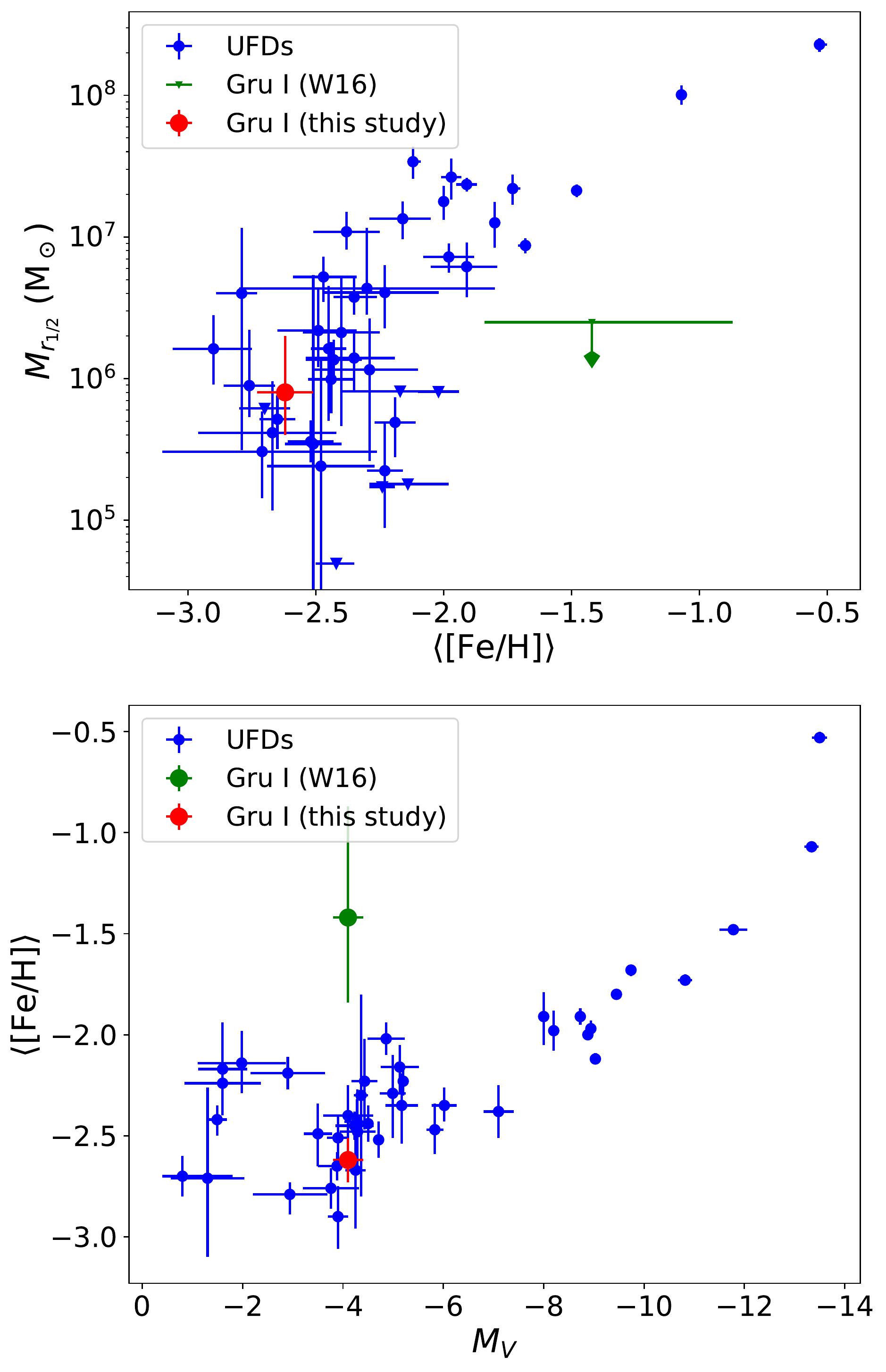}
\caption{Top: UFD dynamical masses as a function of UFD mean metallicity (see Appendix~\ref{app:refs} for a full list of references). 
The Grus~I properties derived in \citet{wmo+16} are shown in green.
The properties derived in this study are shown in red. 
Our derived metallicity and dynamical mass (see Table~\ref{tab:gru1table}) place Gru I in the expected mass-metallicity regime for UFDs.
Bottom: Same as above, but for the metallicity-luminosity relation.
Our results place Grus~I exactly on the UFD trend.}
\label{fig:massmetallicity}
\end{figure}

We derive that Grus~I has a mean metallicity of $\mu_{\text{[Fe/H]}} = -2.62 \pm 0.11$ and place a $2\sigma$ upper-limit on the metallicity dispersion of $\sigma_{\text{[Fe/H]}} < 0.45$ (see right panels in Figure~\ref{fig:cornerplots}). 
The mean metallicity places Grus~I exactly on the mass-metallicity and metallicity-luminosity relation for UFDs (see Figure~\ref{fig:massmetallicity}), affirming its status as a UFD.
The agreement between the location of Grus I on these planes and the population of UFDs also suggests that Grus~I did not experience any unique effects from e.g., tidal stripping relative to the UFD population.
The lack of a resolved metallicity dispersion in Grus I is most likely due to the small sample size of members, for it is not uncommon for UFDs to show metallicity dispersions below our upper limit of 0.45\,dex (see Supplemental Table 1 in \citealt{s+19}).

For completeness, we note that including the metal-rich star at the systemic velocity of Grus I (DES\,J225643.20$-$501130.0; see paragraph 2 in Section~\ref{sec:membership}) returns a mean metallicity of $-2.46^{+0.21}_{-0.19}$\,dex and, unsurprisingly, a resolved metallicity dispersion of $0.53^{+0.25}_{-0.19}$\,dex. 
Consequently, even if we were to assume that DES\,J225643.20$-$501130.0 were a member, Grus I would still have an overall low metallicity in line with other UFDs (see Figure~\ref{fig:massmetallicity}).
The hypothetically resolved metallicity dispersion would also only strengthen the conclusion that Grus I is a UFD. 
High-resolution spectroscopic follow-up of DES\,J225643.20$-$501130.0 is not easily obtainable due to its faintness ($g \sim 19.5$), but would remove any ambiguity in the star's association with Grus~I by allowing neutron-capture element abundances to be derived \citep[e.g.,][]{jsf+19}.

\subsection{J-factor calculations for Grus~I}
\label{sec:jfactor}

We compute the astrophysical component of the dark matter annihilation flux (J-factor) and decay flux (D-factor) following \citet{ps+19}. Briefly, this involves comparing the observed velocity distribution to a theoretical  velocity dispersion from solutions of the spherical Jeans equations \citep[e.g.,][]{GeringerSameth2015ApJ...801...74G, Bonnivard2015MNRAS.453..849B}. In the Jeans modeling, we assume a NFW dark matter profile, a Plummer distribution for the stellar component, and that the stellar anisotropy is constant with radius. For more details see \citet{ps+19}. 

From the combined IMACS and M2FS data set, we compute integrated J-factors of $\log_{10}{J}=16.2_{-0.7}^{+0.8}, ~16.4_{-0.7}^{+0.8},~ 16.5_{-0.7}^{+0.8}$ within solid angles of $0.1\degr,~0.2\degr, ~0.5\degr$ in logarithmic units of GeV$^{2}$~cm$^{-5}$ and compute integrated D-factors of $\log_{10}{D}=16.3\pm0.4, ~16.7\pm0.4,~17.2\pm0.5$ within solid angles of $0.1\degr,~0.2\degr, ~0.5\degr$ in logarithmic units of GeV~cm$^{-2}$.
The J-factor scales as $J\propto \sigma^4/r_{1/2}d^2$ and the low velocity dispersion and large size lead to a small dark matter flux from Grus~I \citep{ps+19}. 
Due to the low dark matter density of Grus~I, the J-factor is quite small compared to other UFDs at similar distances and has one of the lowest J-factors \citep[second to Crater~II][]{Caldwell2017ApJ...839...20C}.  
For reference, the largest J-factors are $\log_{10}{J}\sim19$ for Segue~1 and Reticulum~II \citep{ps+19}. 
If Grus~I were at its orbital pericenter ($\sim 50 ~{\rm kpc}$) the J-factor would increase by a factor of $\sim7$ but Grus~I would still remain as one of the lowest J-factors for MW satellites.  
Grus~I will likely only be useful in stacked analysis for searches for dark matter annihilation.

\subsection{Orbital history \& central density of Grus~I}
\label{sec:orbit}

We initially follow the steps described in Section 4.4 of \citet{sle+20} to model the orbit of Grus~I.
We first derive a systemic proper motion for Grus~I of $\mua = 0.07 \pm 0.05$\,mas\,yr$^{-1}$ and $\mud = -0.25 \pm 0.07$\,mas\,yr$^{-1}$ by taking the inverse-variance weighted average of the \textit{Gaia} EDR3 proper motions \citep{gaia+16, gaia+21} of its eight confirmed members. 
This systemic proper motion agrees exactly with the result from the mixture model approach in \citet{pel+22} and is consistent with the Grus~I proper motion of $\mua = 0.07 \pm 0.05$\,mas\,yr$^{-1}$ and $\mud = -0.27 \pm 0.07$\,mas\,yr$^{-1}$ reported in \citet{btt+22}.
This systemic proper motion, coupled with the radial velocity, distance, R.A., and Decl. in Table~\ref{tab:gru1table} provides 6D phase space information from which to calculate the orbit of Grus~I given a Galactic potential.

As a first pass, we initialized \texttt{orbit} instances in the \texttt{galpy} package \citep{b+15} for Grus~I in a \texttt{MWPotential2014} potential which had been modified to increase the halo mass to $1.6 \times 10^{12}\,M_\odot$ following e.g., \citet{cs+18}.
We note that this potential does not include the effect of the Large Magellanic Cloud (LMC), and so the orbital parameters in this paragraph are shown for comparison purposes only and should not supplant those in e.g., \citealt{pel+22} (see discussion in next paragraph).
We generate 10000 instances of these orbits, sampling from the distance, proper motion, and velocity measurements in Table~\ref{tab:gru1table} as Gaussian distributions, and integrate forwards and backwards for 2\,Gyr.
We derive a pericenter of 20$^{+13}_{-10}$\,kpc, and find that Grus~I will pass its pericenter in $\sim400$\,Myr. 
Additionally, we find that Grus~I has not had a close encounter (within $\sim$10\,kpc) with the LMC, suggesting that it is not an LMC satellite.
This is in agreement with previous studies \citep[e.g.,][]{ksz+18, btt+22}, and our pericenter is consistent with the non-LMC Grus~I pericenter in \citet{pel+22} of $28^{+16}_{-13}$\,kpc.
This suggests that our updated Grus~I radial velocity relative to \citet{wmo+16} should not have a significant effect on the Grus~I orbital parameters.

\citet{pel+22} re-derive the orbital parameters of Grus~I while accounting for the gravitational effects of the LMC. 
Given their inclusion of the LMC, the parameters in that study should supersede the parameters presented in the previous paragraph. 
We note that their assumed systemic Grus~I radial velocity ($-140.5 \pm 2.0$\,km\,s$^{-1}$ from \citealt{wmo+16}) and proper motions ($\mu_\alpha\,\cos\delta = 0.07 \pm 0.05$\,mas\,yr$^{-1}$ and $\mu_\delta = -0.25 \pm 0.07$\,mas\,yr$^{-1}$) are comparable to what is presented Table~\ref{tab:gru1table}.
Indeed, the only parameter that is marginally different is the systemic velocity of Grus~I (by 3\,km\,s$^{-1}$).
The effect of this difference will be negligible compared to uncertainties arising from the proper motion of the system, meaning the orbit results in \citet{pel+22} from including the LMC should be comparable to what one would derive when assuming the Grus~I parameters in this study.
\citet{pel+22} find that the Grus~I orbit pericenter increases to $49^{+27}_{-23}$\,kpc (see Figure 4 in \citealt{pel+22}) when including the effect of the LMC, and that Grus~I is still likely unassociated with the LMC.

Intriguingly, the derived central density of Grus~I from the Jeans modeling in Section~\ref{sec:jfactor} is $\rho_{1/2} \sim 3.5_{-2.1}^{+5.7}  \times 10^7$\,M$_\odot$\,kpc$^{-3}$, among the lowest of UFDs that show no signs of tidal disruption (see Figure~5 in \citealt{pel+22}).
Given the large pericenter of the orbit of Grus~I when including the influence of the LMC, its density is unlikely to be significantly further suppressed on a short timescale by future tidal encounters with the Milky Way.
As increasing samples of UFDs are discovered and characterized with upcoming surveys (e.g., LSST), models of the formation and evolution of UFDs will need to explain this diversity of inner densities independently of mass loss scenarios from interactions with the Milky Way (e.g., \citealt{jkl+21}). 


\subsection{Comparison to previous studies}
\label{sec:comparison}

We derive metallicity and kinematic properties of Grus~I that are more precise than those currently in the literature, largely due to our addition of a comprehensive sample of IMACS velocities and metallicities. 
In particular, we resolve a velocity dispersion ($\sigma = 2.5^{+1.3}_{-0.8}$\,km\,s$^{-1}$), find that Grus~I is a dark-matter-dominated system ($M_{1/2}\,(r_h) =  8^{+12}_{-4} \times 10^5$\,M$_{\odot}$ and M/L$_{V} = 440^{+650}_{-250}$\,M$_{\odot}$/L$_{\odot}$), and that Grus~I has a mean metallicity ($\langle$[Fe/H]$\rangle = -2.62 \pm 0.11$\,dex) that is among the lowest of known UFDs (see Figure~\ref{fig:massmetallicity}). 

Our derived quantities are also generally consistent with the existing upper limits and quantities in the literature. 
\citet{wmo+16} derive a systemic radial velocity of $-140.5^{+2.4}_{-1.6}$\,km\,s$^{-1}$, a dispersion of $\sigma_{v_{\text{los}}} < 9.8$\,km\,s$^{-1}$ and a dynamical mass of $M_{1/2}\,(r_h) < 2.5 \times 10^6$\,M$_{\odot}$, which are consistent with our quantities. 
Their Grus~I mean metallicity ($\langle$[Fe/H]$\rangle = -1.42^{+0.55}_{-0.42}$) is just consistent at the 2$\sigma$ level with our derived metallicity if one also accounts for the +0.32\,dex offset that was applied to M2FS metallicities (see discussion in paragraph 2 of Section~\ref{sec:m2fs}).
\citet{zjb+21} derive a systemic velocity of $-139.2^{+6.1}_{-5.2}$\,km\,s$^{-1}$, a velocity dispersion of $10.4^{+9.3}_{-5.1}$\,km\,s$^{-1}$, and a mass and mass-to-light ratio of $M_{1/2}\,(r_{1/2}) = 1.1^{+2.1}_{-0.8} \times 10^6\,$\,M$_\odot - 1.7^{+2.9}_{-1.2} \times 10^6 $\,M$_\odot$ and $7.4^{+60.2}_{-6.3} \times 10^3$\,M$_\odot$\,L$_\odot^{-1} - 3.2^{+33.9}_{-2.7} \times 10^4$\,M$_\odot$\,L$_\odot^{-1}$, respectively. 
These dispersion, mass, and mass-to-light values are systematically higher than our derived quantities, but still consistent within 2$\sigma$. 
We highlight that our larger sample of members than the previous M2FS study, and our precise IMACS spectroscopic metallicity values allow us to cleanly separate Grus~I members from the foreground.
This leads to more robust constraints on the dynamical properties of the system, through less uncertain kinematic parameters.
We thus conclusively show that Grus~I is a canonical low-metallicity, dark-matter-dominated UFD.

\section{Conclusion}
\label{sec:conclusion}

We present a comprehensive study of the metallicity and kinematic properties of Grus~I, confirming it to be a very metal-poor ($\langle$[Fe/H]$\rangle = -2.62 \pm 0.11$\,dex), dark-matter-dominated (M/L$_{V} = 440^{+650}_{-250}$\,M$_{\odot}$/L$_{\odot}$) UFD.
We combine existing M2FS spectroscopic measurements of Grus~I members in the literature \citep{wmo+16} with comprehensive IMACS spectroscopic follow-up of known and newly discovered members. 
With our updated sample of eight confirmed Grus~I members, we significantly revise downward the existing spectroscopic metallicity of Grus~I ($\langle$[Fe/H]$\rangle = -1.42^{+0.55}_{-0.42}$; \citealt{wmo+16}) and consequently find that Grus~I is one of the lowest metallicity UFDs.
We also resolve a velocity dispersion of $\sigma = 2.5^{+1.3}_{-0.8}$\,km\,s$^{-1}$, consistent with the existing upper limit of $\sigma_{v_{\text{los}}} < 9.8$\,km\,s$^{-1}$ from \citet{wmo+16} and below the previously reported dispersion of $10.4^{+9.3}_{-5.1}$\,km\,s$^{-1}$ in \citet{zjb+21}.
Our analysis corroborates existing hints in the literature from e.g., neutron-capture element abundances \citep{jsf+19} that Grus~I is a UFD.

We note that our additional IMACS observations robustly constrain the properties of Grus~I in two ways.
First, the velocity baseline of our IMACS observations extends from 2015 to 2021 and all of the candidate members in \citet{wmo+16} were re-observed.
This dataset, when coupled with velocities from \citet{wmo+16} and MIKE observations in \citet{jsf+19}, allows us to test nearly every Grus~I member for binarity (see Section~\ref{sec:binaries}).
We identified two Grus~I members as possible binaries and excluded them from subsequent dynamical analysis to avoid biases. 
Second, our IMACS metallicities are more precise than the bulk of existing Grus~I metallicities (see Section~\ref{sec:m2fs}), which allows for a more accurate determination of the metallicity properties of Grus~I and also a cleaner separation of members from the foreground.
From our revised analyses, it is clear that Grus~I is not an anomalous UFD.  We find no evidence of significant mass loss through tidal interactions, and our metallicity measurements place the galaxy on the UFD mass-metallicity and mass-luminosity relations (see Figure~\ref{fig:massmetallicity}).
Table~\ref{tab:gru1table} lists the full properties of Grus~I.

We perform an orbital analysis of Grus~I in a simple Milky Way potential, and find that Grus~I is unlikely to be associated with the LMC. 
As described in \citet{pel+22}, the Grus~I orbital properties are notably affected by the including gravitational effect of the LMC by e.g., shifting its pericenter to $\sim50$\,kpc.
However, those updated parameters still show no evidence that Grus~I is tidally disrupting or associated with the LMC \citep{pel+22}.

Interestingly, the central density of Grus~I ($\rho_{1/2} \sim 3.5_{-2.1}^{+5.7} \times 10^7$\,M$_\odot$\,kpc$^{-3}$) is among the lowest of UFDs that are not known to be tidally disrupting (see Figure~5 in \citealt{pel+22}). 
Only Grus~II has a lower density and Columba~I has an upper limit that is a factor of three larger; however, a number of classical dwarf galaxies (e.g., Ant 2, Crater 2, Fornax) have lower densities.
Despite its low density, it is unlikely that Grus~I is disrupting given its orbit and agreement with UFD scaling relations (see Section~\ref{sec:fehspread}).
Much as models of UFD evolution attempt to explain the diversity in the outskirts of these systems \citep[e.g.,][]{cfs+21, tyf+21}, explaining the variations of their properties in general (e.g., inner densities) will also be key to understanding the evolution of these relic galaxies.
Future surveys (e.g., LSST) have the potential to discover large samples of faint systems in the Milky Way and the Local Group (e.g., \citealt{msc+21}) to assess the full range of their properties.


\begin{deluxetable*}{lllllll} 
\tablecolumns{8}
\tablecaption{\label{tab:gru1feh} IMACS Metallicities of confirmed Grus\,I members}
\tablehead{ID & RA & DEC & $g$\tablenotemark{{\scriptsize a}} & $r$\tablenotemark{{\scriptsize a}} & [Fe/H] &  MEM \\
 & (deg) & (deg) & (mag) & (mag) &  &    }
\startdata
DES\,J225658.06$-$501357.9 & 344.24192 & $-$50.23276 & 18.55 & 17.49 & $-2.66 \pm 0.19$ & M \\
DES\,J225640.78$-$501051.4 & 344.16992 & $-$50.18096 & 20.43 & 19.79 & $-2.09 \pm 0.29$ & M \\
DES\,J225629.92$-$500433.3 & 344.12467 & $-$50.07593 & 19.14 & 18.29 & $-2.45 \pm 0.21$ & M \\
DES\,J225619.67$-$500913.1 & 344.08198 & $-$50.15364 & 20.90 & 20.34 & $-2.79 \pm 0.28$ & M \\
DES\,J225637.05$-$501024.8 & 344.15438 & $-$50.17357 & 20.46 & 19.81 & $-2.59 \pm 0.27$ & M \\
DES\,J225643.29$-$500607.3 & 344.18038 & $-$50.10203 & 19.74 & 19.07 & $-2.90 \pm 0.22$ & M \\
DES\,J225603.76$-$500524.5 & 344.0157 & $-$50.09014 & 20.94 & 20.31 & $-2.70 \pm 0.43$ & M \\
DES\,J225649.23$-$501031.4 & 344.20513 & $-$50.17539 & 20.79 & 20.22 & $-2.92 \pm 0.48$ & M \\
\enddata
\tablenotetext{a}{Quoted magnitudes represent the weighted-average dereddened PSF magnitude derived from the DES DR2 catalog \citep{fdh+15, mgm+18, aaa+21}.}

\end{deluxetable*}

\acknowledgements

A.C. is supported by a Brinson Prize Fellowship at the University of Chicago/KICP.  J.D.S is supported in part by the National Science Foundation under grant AST-1714873.  
A.F. acknowledges support from NSF grant AST-1716251. 
A.B.P. is supported by NSF grant AST-1813881.
T.S.L acknowledges financial support from Natural Sciences and Engineering Research Council of Canada (NSERC) through grant RGPIN-2022-04794.

This work has made use of data from the European Space Agency (ESA) mission
{\it Gaia} (\url{https://www.cosmos.esa.int/gaia}), processed by the {\it Gaia}
Data Processing and Analysis Consortium (DPAC,
\url{https://www.cosmos.esa.int/web/gaia/dpac/consortium}). Funding for the DPAC
has been provided by national institutions, in particular the institutions
participating in the {\it Gaia} Multilateral Agreement.

This project used public archival data from the Dark Energy Survey (DES). Funding for the DES Projects has been provided by the U.S. Department of Energy, the U.S. National Science Foundation, the Ministry of Science and Education of Spain, the Science and Technology FacilitiesCouncil of the United Kingdom, the Higher Education Funding Council for England, the National Center for Supercomputing Applications at the University of Illinois at Urbana-Champaign, the Kavli Institute of Cosmological Physics at the University of Chicago, the Center for Cosmology and Astro-Particle Physics at the Ohio State University, the Mitchell Institute for Fundamental Physics and Astronomy at Texas A\&M University, Financiadora de Estudos e Projetos, Funda{\c c}{\~a}o Carlos Chagas Filho de Amparo {\`a} Pesquisa do Estado do Rio de Janeiro, Conselho Nacional de Desenvolvimento Cient{\'i}fico e Tecnol{\'o}gico and the Minist{\'e}rio da Ci{\^e}ncia, Tecnologia e Inova{\c c}{\~a}o, the Deutsche Forschungsgemeinschaft, and the Collaborating Institutions in the Dark Energy Survey.

The Collaborating Institutions are Argonne National Laboratory, the University of California at Santa Cruz, the University of Cambridge, Centro de Investigaciones Energ{\'e}ticas, Medioambientales y Tecnol{\'o}gicas-Madrid, the University of Chicago, University College London, the DES-Brazil Consortium, the University of Edinburgh, the Eidgen{\"o}ssische Technische Hochschule (ETH) Z{\"u}rich,  Fermi National Accelerator Laboratory, the University of Illinois at Urbana-Champaign, the Institut de Ci{\`e}ncies de l'Espai (IEEC/CSIC), the Institut de F{\'i}sica d'Altes Energies, Lawrence Berkeley National Laboratory, the Ludwig-Maximilians Universit{\"a}t M{\"u}nchen and the associated Excellence Cluster Universe, the University of Michigan, the National Optical Astronomy Observatory, the University of Nottingham, The Ohio State University, the OzDES Membership Consortium, the University of Pennsylvania, the University of Portsmouth, SLAC National Accelerator Laboratory, Stanford University, the University of Sussex, and Texas A\&M University.

Based in part on observations at Cerro Tololo Inter-American Observatory, National Optical Astronomy Observatory, which is operated by the Association of Universities for Research in Astronomy (AURA) under a cooperative agreement with the National Science Foundation.

This research has made use of NASA's Astrophysics Data System Bibliographic Services \citep{woe+00}.

\software{COSMOS reduction pipeline \citep{dbh+11, ock+17} , astropy \citep{astropy, astropy2}, emcee \citep{emcee1, emcee2}, galpy \citep{b+15}}

\appendix

\section{References for dwarf galaxy data in Figure~4}
\label{app:refs}

Here we list the references for the masses, metallicities, and luminosities of the dwarf galaxies plotted in Figure~\ref{fig:massmetallicity}: 
  \citet{msw+03}; \citet{bth+06}; \citet{sg+07}; \citet{bic+08}; \citet{dhc+08}; \citet{mow+08}; \citet{oay+08}; \citet{kwk+09}; \citet{wmo+09b}; \citet{wmo+09}; \citet{sgm+11}; \citet{wgs+11}; \citet{fmt+12}; \citet{kbc+13}; \citet{kcg+13}; \citet{fsk+14}; \citet{bdb+15}; \citet{dbr+15}; \citet{kcb+15}; \citet{kjm+15}; \citet{ksc+15}; \citet{sdl+15}; \citet{csz+16}; \citet{jfs+16}; \citet{kjg+16}; \citet{tkb+16b}; \citet{tkb+16a}; \citet{cts+17};  \citet{cwm+17}; \citet{kcs+17}; \citet{lsd+17}; \citet{mbi+17}; \citet{sld+17}; \citet{smw+17}; \citet{cfj+18}; \citet{kwb+18}; \citet{lms+18}; \citet{lsp+18}; \citet{mcs+18};  \citet{msc+18}; \citet{tbk+18}; \citet{s+19}; \citet{sle+20}; \citet{jlp+21}; \citet{lmi+21}; \citet{cfj+22}; \citet{csl+22};.


\section{Compilation of velocity measurements of Grus I members}
\label{app:vels}

In Table~\ref{tab:literaturevelocities}, we compile all velocity measurements of confirmed Grus~I members in this work with their velocity measurements in \citet{wmo+16}.
Note that a zero-point offset of $-2.6$\,km\,s$^{-1}$ has been applied to the velocities in \citet{wmo+16} to account for an offset between M2FS and IMACS velocities (see Section~\ref{sec:m2fs}).

\startlongtable
\begin{deluxetable*}{lllrrrrrr} 
\tablecolumns{6}
\footnotesize
\tablewidth{\textwidth}
\tablecaption{Compilation of velocity measurements for confirmed Grus I members\label{tab:literaturevelocities}}
\tablehead{   
  \colhead{Name} &
  \colhead{MJD\tablenotemark{a}} &
  \colhead{R.A. (deg)} &
  \colhead{DEC (deg)} & 
  \colhead{Instrument} &
  \colhead{$v_{\text{helio}}$} & 
  \colhead{$v_{\text{err}}$} &
  \colhead{Ref.}\\
\colhead{} &
\colhead{} &
\colhead{(J2000)} &
  \colhead{(J2000)} & 
  \colhead{} & 
  \colhead{(km\,s$^{-1}$)} & 
  \colhead{(km\,s$^{-1}$)}  & 
}
\startdata
DES\,J225658.06$-$501357.9 & 57222.3 & 344.24192 & $-$50.23276 & M2FS & $-141.0$\tablenotemark{b} & $0.4$ & Gru1-032 in \citet{wmo+16} \\
\nodata & 57229.3 & 344.24192 & $-$50.23276 & IMACS & $-140.88$ & 1.11 & This work \\
\nodata & 58762.0 & 344.24192 & $-$50.23276 & IMACS & $-141.59$ & 1.11 & This work \\
\nodata & 59471.0 & 344.24192 & $-$50.23276 & IMACS & $-143.32$ & 1.13 & This work \\
DES\,J225640.78$-$501051.4 & 57222.3 & 344.16992 & $-$50.18096 & M2FS & $-142.0$\tablenotemark{b} & 1.4 & Gru1-004 in \citet{wmo+16} \\
\nodata & 58762.0 & 344.16992 & $-$50.18096 & IMACS & $-142.49$ & 1.46 & This work \\
\nodata & 59471.0 & 344.16992 & $-$50.18096 & IMACS & $-140.68$ & 2.11 & This work \\
DES\,J225629.92$-$500433.3 & 57222.3 & 344.12467 & $-$50.07593 & M2FS & $-146.9$\tablenotemark{b} & 0.8 & Gru1-038 in \citet{wmo+16} \\
\nodata & 57229.3 & 344.12467 & $-$50.07593 & IMACS  & $-147.39$ & 1.15 & This work \\
\nodata & 58762.0 & 344.12467 & $-$50.07593 & IMACS & $-146.38$ & 1.14 & This work \\
\nodata & 59471.0 & 344.12467 & $-$50.07593 & IMACS & $-144.72$ & 1.26 & This work \\
DES\,J225619.67$-$500913.1 & 57222.3 & 344.08198 & $-$50.15364 & M2FS & $-143.8$ &  19.2 & Gru1-022 in \citet{wmo+16} \\
\nodata & 57229.3 & 344.08198 & $-$50.15364 & IMACS & $-146.35$ & 2.39 & This work \\
\nodata & 58762.0 & 344.08198 & $-$50.15364 & IMACS & $-139.67$ & 2.24 & This work \\
DES\,J225637.05$-$501024.8 & 57222.3 & 344.15438 & $-$50.17357 & M2FS & $-140.2$ & 3.9 & Gru1-003 in \citet{wmo+16} \\
\nodata & 57229.3 & 344.15438 & $-$50.17357 & IMACS & $-135.25$ & 2.13 & This work \\
\nodata & 58762.0 & 344.15438 & $-$50.17357 & IMACS & $-143.08$ & 1.58 & This work \\
DES\,J225643.29$-$500607.3 & 57229.3 & 344.18038 & $-$50.10203 & IMACS & $-146.79$ & 1.25 & This work \\
\nodata & 58762.0 & 344.18038 & $-$50.10203 & IMACS & $-146.41$ & 1.24 & This work \\
DES\,J225603.76$-$500524.5 & 57222.3 & 344.01570 & $-$50.09014 & M2FS & $-128.0$ & 87.9 &  Gru1-054 in \citet{wmo+16}\\
\nodata & 58762.03 & 344.0157 & $-$50.09014 & IMACS & $-141.21$ & 1.67 & This work \\
DES\,J225649.23$-$501031.4 & 58762.0 & 344.20513 & $-$50.17539 & IMACS & $-142.47$ & 1.83 & This work \\
\enddata
\tablenotetext{a}{Defined as the MJD at the midpoint of observation. For velocities reported in \citet{wmo+16}, we list the MJD derived from Table 1 in that study.}
\tablenotetext{b}{Offset of $-2.6$\,km\,s$^{-1}$ has been applied to account for a zero-point offset between M2FS and IMACS velocities (see paragraph 3 in Section~\ref{sec:m2fs}).}
\end{deluxetable*}

\bibliography{skymapper}

\end{document}